\pdfoutput=1
\documentclass[12pt,a4paper]{iopart}
\usepackage[utf8]{inputenc}
\expandafter\let\csname equation*\endcsname\relax
\expandafter\let\csname endequation*\endcsname\relax
\usepackage{cite}
\usepackage{amsmath}
\numberwithin{equation}{subsection}
\usepackage{amssymb}
\usepackage{mathtools}
\usepackage{booktabs}
\usepackage{tcolorbox}
\usepackage{enumitem}
\usepackage{doi}
\usepackage{tikz}
\usepackage{hyperref,xcolor}
\hypersetup{colorlinks=true, allcolors=blue}

\usepackage{doi}

\def \fullin{\tikz[baseline=-0.5ex] {
\fill (0,0) circle (1.9pt) coordinate (i);
\node at (0,0)[left]{\footnotesize $i$};
\fill (4ex,0) circle (1.9pt) coordinate (j);
\fill (8ex,0) circle (1.9pt) coordinate (i');
\node at (8ex,0)[right]{\footnotesize $i'$};
\draw (i)--(j);
\draw[densely dashed] (j)--(i');}
}

\def \copiedout{\tikz[baseline=-0.5ex] {
\fill (0,0) circle (1.9pt) coordinate (i);
\node at (0,0)[left]{\footnotesize $i$};
\fill (4ex,0) circle (1.9pt) coordinate (j);
\fill (8ex,0) circle (1.9pt) coordinate (i');
\node at (8ex,0)[right]{\footnotesize $i'$};

\draw (i)--(j);
\draw[densely dashed] (j) (i');}
}

\def \origout{\tikz[baseline=-0.5ex] {
\fill (0,0) circle (1.9pt) coordinate (i);
\node at (0,0)[left]{\footnotesize $i$};
\fill (4ex,0) circle (1.9pt) coordinate (j);
\fill (8ex,0) circle (1.9pt) coordinate (i');
\node at (8ex,0)[right]{\footnotesize $i'$};
\draw (i) (j);
\draw[densely dashed] (j)--(i');}}

\def \bothout{\tikz[baseline=-0.5ex] {
\fill (0,0) circle (1.9pt) coordinate (i);
\node at (0,0)[left]{\footnotesize $i$};
\fill (4ex,0) circle (1.9pt) coordinate (j);
\fill (8ex,0) circle (1.9pt) coordinate (i');
\node at (8ex,0)[right]{\footnotesize $i'$};
\draw (i) (j);
\draw[densely dashed] (j) (i');}}

\def \initg{\tikz[baseline=-0.5ex] {
\fill (0,0) circle (1.9pt) coordinate (i);
\fill (2ex,0) circle (1.9pt) coordinate (j);
\draw (i)--(j);}}

\def \initgpol{\tikz[baseline=-0.5ex] {
\fill (0,0) circle (1.9pt) coordinate (i);
\fill (2ex,0) circle (1.9pt) coordinate (j);}}

\def \initgsolval{\tikz[baseline=-0.5ex] {
\fill (0,0) circle (1.9pt) coordinate (i);
\fill (2ex,0) circle (1.9pt) coordinate (j);
\fill (1ex,1ex) circle (1.9pt) coordinate (k);
\fill (1ex,-1ex) circle (1.9pt) coordinate (l);
\draw (i)--(j);
\draw (i)--(k);
\draw (i)--(l);
\draw (j)--(k);
\draw (j)--(l);
\draw (k)--(l);
}}

\def \initgrtt{\tikz[baseline=-0.5ex] {
\fill (0,0) circle (1.9pt) coordinate (i);
}}

\def \cherry{\tikz[baseline=-0.5ex] {
\fill (0,0) circle (1.4pt) coordinate (i);
\fill (1ex,0.7ex) circle (1.4pt) coordinate (j);
\fill (2ex,0) circle (1.4pt) coordinate (k);
\draw (i)--(j);
\draw (k)--(j);
}}

\def \diamd{\tikz[baseline=-0.5ex] {
\fill (0,0ex) circle (1.4pt) coordinate (i);
\fill (2ex,0ex) circle (1.4pt) coordinate (j);
\fill (1ex,0.7ex) circle (1.4pt) coordinate (k);
\fill (1ex,-0.7ex) circle (1.4pt) coordinate (l);
\draw (i)--(k);
\draw (j)--(k);
\draw (i)--(l);
\draw (k)--(l);
\draw (j)--(l);
}}

\expandafter\def\expandafter\normalsize\expandafter{%
 \normalsize 
 \setlength\abovedisplayskip{0ex}
 \setlength\belowdisplayskip{2.5ex}
 \setlength\abovedisplayshortskip{0ex}
 \setlength\belowdisplayshortskip{2.5ex}
}

\usepackage{iopams} 

\begin{document}

\review[]{Duplication-divergence growing graph models}

\author{Dario Borrelli}

\address{Theoretical Physics Div., University of Naples Federico II, I-80125, Naples, Italy}
\ead{dario.borrelli@unina.it}

\begin{indented}
\item[] (\today)
\end{indented}
\
\begin{abstract}
In recent decades, it has been emphasized that the evolving structure of networks may be shaped by interaction principles that yield sparse graphs with a vertex degree distribution exhibiting an algebraic tail, and other structural traits that are not featured in traditional random graphs. In this respect, through a mean-field approach, this review tackles the statistical physics of graph models based on the interaction principle of duplication-divergence. Additional sophistications extending the duplication-divergence model are also reviewed as well as generalizations of other known models. Possible research gaps and related prior results are then discussed.
\end{abstract}

%
\vspace{0.15in}

\tableofcontents

\title[Duplication-divergence growing graph models
]{\vspace{-0.65in}}

\begin{sloppypar}
\section{Introduction}

What principles underlie the emergence of complexity in large systems represented through the abstraction of graphs? The understanding of the complexity of natural and non-natural systems has often stimulated the intellectual curiosity of humankind. Complexity has been relevant in many physical systems studied over the years, for instance, in relation to collective behaviors in many-body physics and critical phenomena \cite{caianiello1964lectures, coniglio1976percolation,coniglio1977percolation,cowan1994complexity,pines2018emerging}.  The study of complex systems and complexity science has been recognized as a much valuable approach to understand the physical world (see \cite{parisi2023thoughts,bianconi2023complex,Krakauer_FP_24}, and references therein), yet also, complexity is a relevant theme for future research as, e.g., it was argued by theoretical physicist and cosmologist S. Hawking that ``we are approaching the (21st) century of complexity" (quoted in \cite{chui2000unified} pg. 29A).

Among possible definitions of complexity, one that well relates to the context hereafter discussed is given by K. Christensen and N. R. Moloney \cite{christensen2005complexity} in their \textit{Complexity and Criticality}---`\textit{the repeated realization of simple principles in systems with many degrees of freedom that gives rise to emergent behavior not encoded in principles themselves.}'

The systems mentioned by this definition may be known as `complex systems', i.e., ensembles of many interacting constituents that according to basic principles of interaction give rise to collective behavior that may not be understood by the only study of the individual behavior of each constituent; a concept reminiscent of \textit{More is Different} by P. W. Anderson introduced in \cite{anderson1972more}.

While a universal definition of a complex system (as of complexity) is debated \cite{estrada2023complex}, the emergence of behaviors, patterns, and regularities from ensembles of interacting constituents \cite{flack2011challenges} is often suggested as an hallmark of complex systems, being it consistently recurrent in the scientific literature (e.g., \cite{anderson1972more} in 1972, \cite{strogatz2022fifty} in 2022, \cite{de2023more} in 2023). Statistical physics has often tackled the problem of describing macroscopic behaviors in systems with many degrees of freedom from the interaction of their many-body constituents \cite{caianiello1964lectures,coniglio1976percolation,coniglio1977percolation} (also via the renormalization group theory \cite{wilson1971renormalization,goldenfeld2018lectures,caldarelli2024laplacian,gabrielli2025network}), relating microscopic principles with macroscopic observations.

Within this paradigm, an approach to model complex systems as evolving random graphs has its basis in (non-equilibrium) statistical physics \cite{krapivsky2001organization,albert2002statistical,dorogovtsev2003evolution}. By analogical reasoning, vertices of an evolving graph (i.e., evolving network) can be an abstraction of constituents of a complex system, while interactions between these constituents can be abstracted by edges of a graph. Interaction principles among vertices as well as the emergence of structural properties (e.g., absence of a characteristic scale of observation for some observables) of the whole graph from local interactions are the specific focus of this review. In particular, the interaction principle considered here concerns how interactions among vertices emerge within a graph undergoing growth by \textit{duplication} of existing linkages, and \textit{divergence} by probabilistic changes of these linkages.
Intrinsic features of each vertex are kept aside along with some microscopic degrees of freedom that observable systems may present, favoring an essential approach to modeling, typical of theoretical physics.  Specifically, the aim of this paper is to review published studies that propose growing network models in which interactions between vertices of evolving graphs arise from duplication (and divergence). These graph models are of a broad interest, e.g., biological networks \cite{sole2002model,kim2002infinite,pastor2003evolving}, scientific citation graphs \cite{krapivsky2005network}, web graphs \cite{kumar2000stochastic}, online social graphs \cite{bhat2016densification}. Also, they are known to be among possible explanations for the emergence of preferential-attachment in growing networks \cite{dorogovtsev2003evolution}. Since no published review paper is specifically focused on reviewing duplication (and divergence) growing graph models, this has fostered motivation to write this review.\footnote{The following naming conventions are used: Ref.  (Reference); Sec. (Section); Eq. (Equation); Fig. (Figure); $\sim$ (of the order of); $\propto$ (proportional to); $\approx$ or $\simeq$ (approximately equal to); lhs (left hand side of an equation); rhs (right hand side of an equation). Note that when referring to \textit{duplication (and divergence)}, it might include additional sophistications to the simpler \textit{duplication-divergence} principle.}

\subsection{Context}

Scholars in the physical and mathematical sciences made substantial efforts to advance the understanding of networks through the development of \textit{graph theory} (namely, network theory).
These scholars originally referred to the term \textit{random graphs} when treating geometric entities that relate to static (equilibrium) networks with a Poisson distribution of the number of edges per vertex \cite{solomonoff1951connectivity,
gilbert1959random,renyi1959random}; these random graph models were then called Erd{\H{o}}s-R{\'e}nyi random graphs. Similarly, random graphs were introduced earlier (in the 1940s) in the context of statistical physics (specifically, polymer physics) by P. J. Flory and W. H. Stockmayer  \cite{flory1941molecular,stockmayer1943theory}. For many years, this class of random graphs was the typical model of a complex network one could refer to.
Approaching the end of the 20th century, with support of empirical data, it has been manifestly shown that networks are everywhere, whose interest spans many academic fields, leading to the establishment of an interdisciplinary science, namely, network science \cite{barabasi2013network,vespignani2018twenty}.  Moreover, many empirical networks that were studied, while being microscopically different (meaning that constituents of two different networks are different), have shown emergent macroscopic regularities that were surprisingly shared by diverse empirical networks. Among results concerning these emergent features (for a review see, e.g.,  Ref. \cite{albert2002statistical}) one can mention: an algebraic tail in the distribution of the number of edges per vertex (namely, the vertex degree distribution) \cite{barabasi2009scale}, a short average topological distance between any two constituents compared to Erd{\H{o}}s-R{\'e}nyi random graphs of the same graph order  \cite{watts1998collective}, hidden metric spaces able to explain the underlying geometry of networks \cite{boguna2021network}, networks of higher-order such as hypergraphs and simplicial complexes \cite{bianconi2021higher}.  Many features reflecting the structure of empirical networks were not expected in random graphs with a Poisson distribution of the number of edges per vertex, and therefore, new graph models were required to describe emergent behaviors as well as new approaches to study random graphs \cite{durrett2010random,van2024random}. A crucial observation is that many networks are typically dynamic \cite{holme2012temporal,masuda2016guide}, as they evolve with time or they change in response to stimuli, with their non-equilibrium characteristics that may not be understood without studying principles of their growth and evolution \cite{dorogovtsev2003evolution}.

Based on (non-equilibrium) statistical physics and graph theory, for at least two decades, a research effort has been made to enhance the understanding of random graph models to describe heterogeneous and evolving networks \cite{dorogovtsev2003evolution}. The aim of this review is to provide an introduction to this research effort with respect to a peculiar type of growing network models known as \textit{duplication (and divergence)} models. Then, it would be possible to identify ideas for future research that arise from reviewing prior models.

\subsection{Models of growing graphs}

Growing graph models (or, equivalently, network growth models) are based on a network generation process which iteratively adds new vertices and edges. A new vertex attaches to a constant or a variable number of existing vertices chosen either deterministically or stochastically. The procedure by which new edges and new vertices join the growing graph distinguishes between specific types of growing graph models. When there is any random variable involved in the growing process, the model may be called a \textit{random growing graph} model \cite{dorogovtsev2003evolution,durrett2010random}. In the statistical physics sense, a random growing graph may be considered as an ensemble of realizations of growing graphs. In random growing graph models here considered, the number of vertices and edges of a graph increases depending on random variables. In the specific case in which a single vertex is added at each iteration, a random growing graph model is called a \textit{sequentially growing random graph} model.

The term `sequentially' denotes that one vertex at a time is added during growth. This particular condition, when possible, allows one to write an identity between the number of vertices in the graph and the number of iterations (i.e., a discrete time variable) of the growth process. Then, if one considers a continuum approximation, this assumption may often allow to differentiate, or to integrate, with respect to the number of vertices, or with respect to time, indistinguishably. As it will be shown, prior work leveraged this approximation to derive analytic forms of the evolution of observable quantities of sequentially growing random graphs. In other cases discussed here, it can happen that, by model construction, the aforementioned identity between time and the number of vertices does not hold because the probability of new vertices joining the graph is less than 1 per iteration.

Vertices joining the growing network by attaching to one randomly chosen existing vertex allow one to define the simplest sequentially growing random graph model, i.e., the \textit{random recursive tree}. Yet, there are different ways a new vertex can join existing vertices of a growing graph. For instance, when a single added vertex choses an existing vertex to connect to with a probability proportional to its number of edges, it reflects a situation of network growth by \textit{preferential-attachment}. In other words, the attachment rate of a new vertex to existing vertices is proportional to their vertex degree (i.e., number of edges). In case of linear proportionality, such a sequentially growing random graph model is called \textit{linear preferential-attachment} model, also known as the Barabási-Albert model \cite{barabasi1999mean}. Instead, when the attachment rate is a nonlinear function of the vertex degree of existing vertices the model is typically referred to as a \textit{nonlinear preferential-attachment} model \cite{krapivsky2010kinetic,zadorozhnyi2015growing}.

\subsection{Duplication (and divergence) growing graphs}
Among the many diverse cases of sequentially growing random graphs (e.g., by preferential-attachment), a new vertex may instead arise from duplication of an existing vertex (and of its edges).

The new vertex can choose an existing vertex to copy its edges from, without any preferential selection, choosing a vertex among the existing ones in a random uniform choice. This situation leads to the definition of a class of sequentially growing random graph models known as \textit{duplication (and divergence)}, which is the main focus of this review.

While being preferential-attachment and duplication (and divergence) two different approaches to model network growth, it turns out that duplication (and divergence) may be seen as a generative principle for preferential-attachment \cite{vazquez2003growing,dorogovtsev2003evolution}. Indeed, as in the case of preferential-attachment, the rate equation for the vertex degree distribution of duplication (and divergence) model graphs exhibits an attachment rate that is proportional to the vertex degree \cite{kim2002infinite}, particularly similar to linear preferential-attachment \cite{barabasi1999mean}. Whilst there are strong conceptual and analytic links between duplication (and divergence) and preferential-attachment that will be discussed, this review is not primarily concerned with preferential-attachment, for which there is a vast literature available (e.g., see \cite{barabasi1999mean,dorogovtsev2000structure,krapivsky2001organization,albert2002statistical,dorogovtsev2003evolution} and references therein).

In duplication (and divergence) growing graph models, when only a fraction of edges of a randomly chosen vertex is duplicated, duplication models are called \textit{duplication-divergence} models (or \textit{partial duplication} models). Different versions of duplication (and divergence) models include addition of edges (other than those edges that are duplicated) between the added vertex and the copied vertex, or between the added vertex and other existing vertices. Within duplication (and divergence) models there are cases in which the vertex chosen randomly for duplication may lose edges subsequent to duplication in a way that is independent or dependent on conservation or loss of duplicate edges by the copy vertex. Before reviewing duplication (and divergence) models, it is convenient to set some notation and to outline some basic definitions.  This preliminary information would help readability of analytic derivations for models reviewed in subsequent sections.

\subsubsection{Notation and definitions}

A realization of a growing (non-equilibrium) graph at time $t$ is denoted by $G_t = (V_t,E_t)$, where $V_t$ denotes the set of vertices at $t$, and $E_t$ the set of edges at $t$.
A random growing graph $\mathcal{G}_t$ can be considered as a statistical ensemble of realizations $G_t$, each occurring with a certain probability $P_{G_{t}}$ \cite{dorogovtsev2022nature}. When studying an observable $x_t$ of $\mathcal{G}_t$, one can denote moments of its distribution by $\langle x_t \rangle$, $\langle x_t^2 \rangle$, ..., $\langle x_t^n \rangle$, ..., where the symbol $\langle \cdot \rangle$ denotes average over the ensemble.

It is worth noting that there might be cases in which a duplication (and divergence) growing graph model is deterministic (either intrinsically by model construction, or due to a certain value of a parameter). For instance, for same boundary conditions, the mean of an observable $x_t$ of $\mathcal{G}_t$ over the ensemble coincides with its value for a single realization, e.g., the first moment is $\langle x_t \rangle = x_t$ as each realization would still give $x_t$.

\subsubsection{Vertices, probabilities, observables.}

Two vertices are usually the focus of a duplication-divergence iteration: the vertex that is duplicated, namely the \textit{original vertex}, and the vertex that duplicates the original vertex, namely the \textit{copy vertex}. Hereafter, the \textit{original vertex} is always denoted by $i$, and the \textit{copy vertex} by $i'$.

Subsequent to duplication, an edge that has a respective duplicate may be retained or may be lost depending on the specific model considered. The probability of conserving a duplicate edge is typically denoted by $p$, thus, the probability of not conserving a duplicate edge is denoted by $1-p=\delta$; it may assume a different form for each specific model.

When duplicate edges are lost due to divergence, and remaining edges that were duplicated have then one of the two edge ends that is always the copy vertex, or always the original vertex, then the other vertex becomes one with no edges, namely a `\textit{non-interacting}' vertex, (sometimes it is called a `\textit{singleton}', an `\textit{isolated vertex}', which are considered as synonyms of a vertex with no edges).

To set a notation concerning the addition of edges other than those that are duplicated in a \textit{duplication (and divergence)} model iteration, it is worth mentioning that there may be two possibilities: (i) linking the copy vertex to the original vertex, and (ii) linking the copy vertex to vertices other than the adjacent vertices of the original vertex. In case (i), called \textit{dimerization}, the probability of linking the copy vertex to the original vertex is denoted by $\alpha$, whilst, in case (ii), called \textit{mutation}, the probability of linking duplicate vertices to other vertices (except from adjacent vertices of the original vertex) is denoted by $\beta$. New edges are typically emanated from the copy vertex except from cases that will be explicitly mentioned, where new edges arise among prior existing vertices. Note that, as in prior work, $\delta,\alpha,\beta$ may be referred to as rates as well as probabilities (since they are typically defined in $[0,1]$).

An observable worth of mentioning is the vertex degree $k$ of a vertex $j \in V_t$. This quantity represents the number of edge ends attached to a vertex $j $. Because duplication (and divergence) growing graph models here reviewed typically reflect undirected graphs, all edges hereafter considered are non-directed edges (unless otherwise specified). Also, using the nomenclature of \cite{dorogovtsev2003evolution}, graphs considered hereafter have no \textit{tadpoles} (self-edges) and no \textit{melons} (multi-edges). When the graph has $N$ vertices, the number of vertices with vertex degree $k$ is denoted by $N_k(N)$, and often, it is normalized by $N$ giving the fraction of vertices with vertex degree $k$ when the graph has $N$ vertices, i.e., $N_k/N$. From these quantities, one can define the expected fraction of vertices with degree $k$ (namely, the vertex degree distribution) as $n_k = \langle N_k/N\rangle$. Then, $l$th-moments of the vertex degree distribution read respectively as $\langle k \rangle = \sum_k k n_k$, $\langle k^2 \rangle = \sum_k k^2 n_k$, $\dots$, $\langle k^l \rangle = \sum_k k^l n_k$. Other quantities of interest that may be useful to understand results of published models, when encountered, will be properly defined.

\subsubsection{Initial graphs, connected graphs.}
\label{sec_initcondit}

In principle, the initial graph to start a duplication (and divergence) growth is arbitrary. Yet, a minimal initial graph may be a connected graph with two vertices and one edge linking them. We recall that a \textit{connected graph} is a graph in which for any pair of vertices there exists a path joining them;\footnote{Also, a \textit{path} is a walk in which all vertices are distinct. A \textit{walk} is a sequence beginning and ending with vertices, which in between alternates edges and vertices.} this review article refers to this particular choice (i.e., two connected vertices) for the typical initial graph.  Different choices of initial graphs, when considered, will be mentioned.

\subsection{Mean-field approach, and limiting quantities}

Mean-field theory, or mean-field approximation, originates from statistical physics and it is frequently used in descriptions of varying observables of the structure of growing graphs \cite{barabasi1999mean,cai2015mean}. The idea underlying mean-field theory in the context of graphs is to replace all interactions of a vertex with a single mean interaction by an imaginary external field. This review relies on this description to obtain analytic forms describing quantities such as mean vertex degree, mean number of edges, expected vertex degree distribution. The advantage of the mean-field approach is giving a first approximation which is usually very informative and useful as one can get an analytical understanding of the mean behavior of observables of interest in ensembles of random growing graphs.

Moreover, this review presents prior results in analytic forms considering the long-time, large-graph-order limit ($t \rightarrow \infty$, or $N \rightarrow \infty$). This assumption allows one to get a physical sense of the scaling with $t$ (or with the number of vertices $N$) of a certain observable that, on the one hand, may be preferable to exact solutions since exact forms often fail to convey a physical understanding; on the other hand it is a valuable approach to highlight similarities across different duplication (and divergence) growing graph models. While, in some cases, mean-field theory well approximates exact solutions (for instance, the vertex degree distribution in the full duplication graph model), in other cases, exact solutions have not been found yet, and mean-field theory may be the most immediate approach one can currently rely on to get analytic forms of some quantities of interest. In the latter case, we will mainly rely on mean-field approximations while pointing out references to other kinds of solution when available.

\section{Models}

In the following sections, prior published models of duplication (and divergence) are reviewed, with the inclusion of additional sophistications such as dimerization, mutation, and deletion. Models are organized by sorting them via the number of parameters and procedural steps required to achieve a generic growth iteration. Hence, models are introduced beginning with the full duplication model, and then tackling various duplication-divergence models with additional sophistications (e.g., mutation, dimerization, deletion).

The aforementioned sorting of prior models facilitates readability and derivation of quantities (e.g., the mean number of edges, vertex degree, and vertex degree distribution, which are the quantities mainly treated hereafter) which typically extend derivation of less sophisticated models.

\subsection{Full duplication}
\label{sec_fulldup}

\begin{figure}[!t] \vspace{0.2in}
\hspace{2.5cm}
	\includegraphics[width=0.285\textwidth]{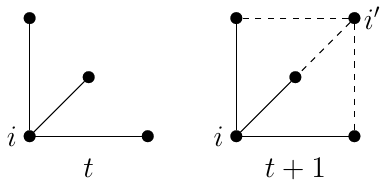}
	\caption{Simplified depiction of a possible realization of an iteration ($t \rightarrow t+1$) of the full duplication model. Dashed edges are duplicated from the original vertex $i$ (chosen uniformly at random) and attached to the copy vertex $i'$.}
	\label{full_dd_fig}
\end{figure}

The first sequentially growing network model reviewed here is based uniquely on the principle of duplication, a model often called the \textit{full duplication} model. The generic iteration of the full duplication model can be formalized by the following procedure (see also Fig.~\ref{full_dd_fig} for a simplified depiction of a realization of a generic iteration):

\begin{tcolorbox}[boxrule=0.8pt, colframe=white, colback=white, sharp corners]
\begin{enumerate}[leftmargin=30pt,itemsep=-0.01in]
	\item Random uniform choice of a vertex $i$ among existing vertices.
	\item Duplication of vertex $i$ into vertex $i'$ with the same set of edges of $i$.
\end{enumerate}
\end{tcolorbox}

Despite appearing as a minimal growth principle, the full duplication model has been widely studied due to its intriguing phenomenology (see \cite{kumar2000stochastic} for one of the earliest references). As shown subsequently, more articulate models with the divergence process, in special cases, can reproduce the full duplication model.

\subsubsection{Mean vertex degree.}

To write an expression for the mean vertex degree for sequentially growing graphs by full duplication one can begin by considering the number of edges at time $t$, i.e., $E_t := |E_t|$, averaged over the ensemble of realizations, i.e., $\langle E_t \rangle$. Then, one can write

\begin{equation}
	\langle E_{t+1} \rangle  = \langle E_t \rangle + \langle k_t \rangle,
\end{equation}

because at each iteration, in the spirit of the mean-field approach, a number $\langle k_t \rangle$  of edges are added by duplication of a vertex chosen uniformly at random among the set of existing vertices at time $t$.
Knowing that $\langle k_t \rangle = 2\langle E_t \rangle/t$, the above equation is rewritten as

\begin{equation}
	(t+1) \langle k_{t+1} \rangle = t \langle k_t \rangle + 2\langle k_t \rangle.
\end{equation}

Solving this difference equation, gives the following scaling with $t$

\begin{equation}
	\langle k_{t} \rangle \sim  t,
	\label{fdd}
\end{equation}

suggesting that in the full duplication model the mean vertex degree might scale linearly with $t$.

\subsubsection{Mean number of edges.}

Any realization of the full duplication model starting with two connected vertices (which is the initial graph $G_{t_0=2}$ hereafter considered, as mentioned in Sec.~\ref{sec_initcondit}) is a complete bipartite graph \cite{kim2002infinite,ispolatov2005duplication}, i.e., a graph with two distinct subsets of vertices with each vertex of one subset connected to all vertices of the other subset (and vice versa).

Thus, a realization of a full duplication model with initial graph $G_{t_0=2}$ having two vertices connected by an edge is a complete bipartite graph $\mathcal{K}_{j,t-j}$ \cite{ispolatov2005duplication}, with one subset of $j$ vertices and the other subset of $t-j$ vertices, with $j=1,...,t-1$, occurring equiprobably.

The total number of edges of $\mathcal{K}_{j,t-j}$ is $E_t = j(t-j)$. Averaging over the ensemble of realizations one can get $l\text{th}$-moments of the distribution (namely, $\langle E_t^l \rangle$). For instance, the first moment ($l=1$) is

\begin{equation}
\langle E_t \rangle = \frac{1}{t-1}\sum_{j=1}^{N-1} j(t-j) = \frac{t(t+1)}{6},
\label{E_t}
\end{equation}
and the second moment ($l=2$) is

\begin{equation}
\langle E_t^2 \rangle = \frac{t(t+1)(t^2 + 1)}{30}.
\label{E_t2}
\end{equation}

Similarly, one can continue for $l>2$ \cite{ispolatov2005cliques}. Moreover, (\ref{E_t}) and (\ref{E_t2}) can be leveraged to check whether or not the random graph considered here is self-averaging, i.e., $\lim_{t \rightarrow \infty} \langle E_t^2 \rangle / \langle E_t \rangle ^2 = 1$. This limit, calculated with (\ref{E_t}) and (\ref{E_t2}), returns $6/5 \neq 1$. Despite the apparent simplicity of model construction, full duplication random graphs exhibit a lack of self-averaging \cite{ispolatov2005duplication}, a very peculiar feature of the full duplication model further deepened within the discussion section.

\subsubsection{Vertex degree distribution.}

The vertex degree distribution of the full duplication model with an initial graph of two connected vertices ($G_{t_0=2}$:\hspace{0.05in}\initg ) can be derived by considering the number of vertices with degree $k$ in the $\mathcal{K}_{j,t-j}$ graph introduced in the previous subsection, i.e.,

\begin{equation}
	 N_k(j) = \delta_{k,j}(t-j) + \delta_{k,t-j}(j),
\end{equation}

where $\delta_{a,b}$ is the Kronecker delta function. When one considers averaging over all realizations, i.e., $(t-1)^{-1}\sum_{j=1}^{t-1}N_k(j)$, and dividing by $t$ (i.e., here equivalently dividing by $N$) as in \cite{kim2002infinite}, it yields

\begin{equation}
	n_k = \frac{2(t-k)}{t(t-1)},
	\label{fulldelt}
\end{equation}

representing, for the case considered here, the distribution of the fraction of vertices with vertex degree $k$, in agreement with Ref.~\cite{ispolatov2005duplication}. As pointed out in \cite{kim2002infinite,ispolatov2005duplication}, individual realizations lead to widely different outcomes, and (\ref{fulldelt}) holds only for the special initial graph having two vertices joined by one edge. An exact solution for the vertex degree distribution of the full duplication with arbitrary initial graph was given in \cite{raval2003some}.

\subsection{Duplication-divergence}
\label{sec_partialdup}
\begin{figure}[!t] \vspace{0.2in}
	\hspace{2.5cm}
	\includegraphics[width=0.58\textwidth]{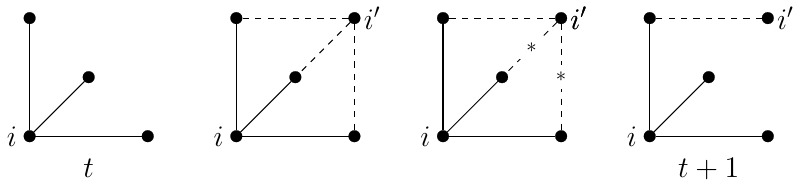}
	\caption{Simplified depiction of a possible realization of an iteration ($t \rightarrow t+1$) of the duplication-divergence model. Dashed edges attached to $i'$ have been duplicated from vertex $i$. Edges marked with $*$ are not conserved due to the divergence process.}
	\label{dd_as_fig}
\end{figure}

As an immediate extension of the full duplication model, this model includes a partial conservation of duplicate edges. The duplication-divergence model is, indeed, also known as the \textit{partial duplication} model. The generic iteration for this model reads as (see also Fig.~\ref{dd_as_fig}):

\begin{tcolorbox}[boxrule=0.8pt, colframe=white, colback=white, sharp corners]
\begin{enumerate}[leftmargin=30pt,itemsep=-0.01in]
	\item Choice of a vertex $i$ uniformly at random among existing vertices.
	\item Duplication of vertex $i$ into vertex $i'$, having  same edges of $i$.
	\item Each duplicate edge $(i',j)$ is conserved with probability $p$ (namely, lost with probability $\delta=1-p$).
\end{enumerate}
\end{tcolorbox}

Note that this process becomes a full duplication (Sec. \ref{sec_fulldup}) when $p=1$ ($\delta=0$); $\delta \in [0,1]$ is called the \textit{divergence rate}. The divergence process described in (iii) is a \textit{complete asymmetric divergence}, as it affects only edges emanated from the copy vertex $i'$, while vertex $i$ conserves all of its edges.

\subsubsection{Mean vertex degree (with non-interacting vertices).}

As in the case of full duplication, and with an additional term accounting for the divergence process, we can write the following equation describing the mean change in the number of edges as

\begin{equation}
	\langle E_{t+1} \rangle - \langle E_t \rangle = \langle k_t \rangle - \delta \langle k_t \rangle.
	\label{partdup_begin}
\end{equation}

Recalling $\langle k_t \rangle = 2\langle E_t \rangle /t$, it becomes

\begin{equation}
	(t+1)\langle k_{t+1} \rangle = t \langle k_t \rangle + 2 \langle k_r \rangle - 2\delta \langle k \rangle,
\end{equation}

which it is recast as

\begin{equation}
	(t+1)\langle k_{t+1} \rangle = (t+1) \langle k_t \rangle +  \langle k_r \rangle - 2\delta \langle k \rangle.
\end{equation}

Solving this recurrence gives the following scaling with $t$

\begin{equation}
	\langle k_t \rangle \sim t^{1-2\delta}.
	\label{sing}
\end{equation}

For $\delta=0$, (\ref{sing}) suggests a linear scaling as in the full duplication model, as it has been shown through (\ref{fdd}).

\subsubsection{Mean vertex degree (without non-interacting vertices).}
When a vertex with degree $k$ is chosen for duplication, and all the duplicate edges are subsequently lost in the divergence process with probability $\varphi = (1-p)^{k}=\delta^k$, the copy vertex becomes a non-interacting vertex and, as in Ref. \cite{ispolatov2005duplication}, it may be removed from the graph, with the resulting graph at $t+1$ unvaried from the graph at $t$. In this case, the mean increment of vertices is therefore

\begin{equation}
	\nu = \sum_k n_k (1-\varphi),
\label{nu_eq}
\end{equation}

and the mean vertex degree scales with $N$ as

\begin{equation}
	\langle k_N \rangle \sim N^{(2p/\nu) -1}.
	\label{nsing}
\end{equation}

Here, $N$ is the number of vertices with at least one edge. Thus, here one can notice the identity $t=N$ is not a valid assumption except if one sets $\nu=1$, which would manifestly turn (\ref{nsing}) into (\ref{sing}).

\subsubsection{Mean number of edges (with non-interacting vertices).}

A recurrence for the mean number of edges can be written as

\begin{equation}
	\langle E_{t+1} \rangle = \langle E_t \rangle \left[ 1 + \frac{2(1-\delta)}{t}\right].
\end{equation}

This recurrence can be solved exactly, e.g., for an initial graph with two connected vertices, yielding

\begin{equation}
	\langle E_t \rangle = \frac{\Gamma(2-2\delta + t)}{\Gamma(t)\Gamma(4-2\delta)},
	\label{extsol}
\end{equation}

with $\Gamma(\cdot)$ the Euler's Gamma function. Eq.~(\ref{extsol}) has been provided in \cite{borrelli2024divergence} within a generalization of duplication-divergence models (see Sec.~\ref{genrl}).

\subsubsection{Mean number of edges (without non-interacting vertices).}

From Ref.~\cite{ispolatov2005duplication}, the mean increment of the number of edges is

\begin{equation}
	\Delta E = \sum_{k=1}^{\infty} n_kpk = p\sum_{k=1}^{\infty} n_k k = p\langle k_N \rangle.
\end{equation}

As it was carried out in \cite{ispolatov2005duplication}, the ratio $\Delta L / \Delta N$ and $\langle k \rangle = 2\langle E_N \rangle /N$ help to write a continuum form in the large $N$ limit

\begin{equation}
\frac{d \langle E_N \rangle }{dN} = \frac{2p}{\nu}\frac{\langle E_N \rangle}{N},
\label{meanedge_ispo}
\end{equation}

which represents the rate equation for the mean number of edges; $\nu$ is from Eq.~(\ref{nu_eq}). When solving (\ref{meanedge_ispo}), it yields the scaling

\begin{equation}
\langle E_N \rangle \sim N^{2p/\nu},
\end{equation}

reminiscent of (\ref{nsing}) when divided by $N$.

\subsubsection{Vertex degree distribution.}
\label{subsec_ispo_degdist}

The number of vertices with degree $k$ for the partial duplication model without non-interacting vertices obeys the following rate equation \cite{ispolatov2005duplication}

\begin{equation}
\nu \frac{dN_k}{dN} = (1-\delta) \left[ (k-1)\frac{N_{k-1}}{N} - k \frac{N_k}{N} \right] + \mathcal{M}_k,
\label{mastdeg1}
\end{equation}

with $\mathcal{M}_k$ given by the following identity

\begin{equation}
\mathcal{M}_k = \sum_{s\geq k} n_s \binom{s}{k}(1-\delta)^{k}\delta^{s-k},
\label{Gk_ispo}
\end{equation}

which, as a result by \cite{kim2002infinite}, it can be approximated by noticing the summand as sharply peaked around $s \approx k/p$, yielding

\begin{equation}
\mathcal{M}_k \approx n_{k/p} (1-\delta)^{-1},
\end{equation}

since $\sum_{s \geq k}\binom{s}{k}(1-\delta)^{k}\delta^{s-k}=(1-\delta)^{-1}$. Eq.~(\ref{mastdeg1}) can be then rewritten as

\begin{equation}
\nu \frac{dN_k}{dN} + (1-\delta)\frac{d(n_k k)}{dk} = n_{k/p} (1-\delta)^{-1}.
\label{remi}
\end{equation}

In \cite{ispolatov2005duplication}, for the partial duplication model (without non-interacting vertices), through scaling assumptions while mixing analytic and numerical results, the following scaling for $n_k$ is written

\begin{equation}
n_k \sim
\begin{cases}
N^{1-2p}f(k/N^{2p-1}), \hspace{1.2cm} \text{for \; } 1/2<p<1,\\
k^{-2}, \hspace{3.7cm} \text{for \; } p^*<p<1/2 ,\\
k^{-\gamma (p)}, \hspace{3.3cm} \text{for \; } 0<p<p^*,
\end{cases}
\label{degdist_cases}
\end{equation}

with $p^*$ estimated to be $e^{-1}$, and $f(\cdot)$ a scaling function. From \cite{ispolatov2005duplication}, the relation between the exponent $\gamma$ and $p$ reads as

\begin{equation}
\gamma (p) = 3 - p^{\gamma - 2}.
\label{gamma_ispo}
\end{equation}

Noteworthy, if one would not neglect non-interacting vertices, then the rate of vertices joining the growing graph would be equal to $1$. However, the rate $\nu$ which appears in Eq.~(\ref{mastdeg1}), as in the case without non-interacting vertices, turns out to be the rate of vertices joining the set of vertices with at least one link (not the set of non-interacting vertices).

When calculating vertex degree distributions one typically assumes $k>0$, therefore, we would naturally return to (\ref{degdist_cases}) by slightly changing the duplication principle in the duplication-divergence process in the following way: ``(i)~Choice of a vertex $i$ uniformly at random among existing vertices with at least one edge" with subsequent (ii), (iii) unvaried from Sec.~\ref{sec_partialdup}.

\subsection{Duplication-divergence with mutation}
\label{sec_dupdivmut}

The full duplication model as well as the duplication-divergence model are growth processes based on reuse of existing patterns of linkage among vertices. In the case of full duplication (Sec.~\ref{sec_fulldup}), a set of edges is exactly duplicated at each iteration, while in the duplication-divergence model (Sec.~\ref{sec_partialdup}), the duplication of edges of the original vertex is only partial.

In this section, duplication and complete asymmetric divergence is accompanied by mutation, which is a sophistication consisting in edges added among the copy vertex and other vertices of the graph (except from vertices adjacent to the original vertex before divergence) \cite{sole2002model}.

Inspired by the context of biomolecular networks \cite{pastor2003evolving}, mutation might include cases in which the copy vertex develops new interactions as it allows the addition of edges with other vertices that are not adjcent to the original vertex. The generic iteration for graphs growing by duplication-divergence with mutation is (see Fig.~\ref{dd_as_m_fig}):

\begin{figure}[t!] \vspace{0.2in}
	\hspace{2.5cm}
	\includegraphics[width=0.73\textwidth]{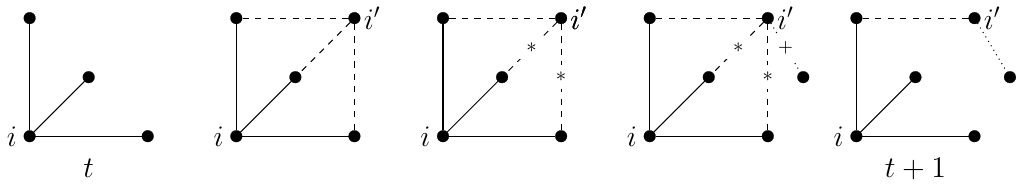}
	\caption{Simplified depiction of a possible realization of an iteration of the growing graph model by duplication-divergence with mutation. Dashed edges attached to $i'$ have been duplicated from vertex $i$. Edges marked with $*$ are not conserved at $t+1$ due to the divergence process. The dotted edge marked with $+$ is added due to mutation.}
	\label{dd_as_m_fig}
\end{figure}

\begin{tcolorbox}[boxrule=0.8pt, colframe=white, colback=white, sharp corners]
\begin{enumerate}[leftmargin=30pt,itemsep=-0.01in]
	\item Choice of a vertex $i$ uniformly at random among existing vertices.
	\item Duplication of vertex $i$ into a vertex $i'$, having the same edges of $i$.
	\item Each duplicate edge $(i',j)$ is conserved with probability $p$ (namely, lost with probability $1-p=\delta$).
	\item An edge between vertex $i'$ and any other vertex of the graph (except from $i$'s adjacent vertices) is added with probability $\beta$.
\end{enumerate}
\end{tcolorbox}

In (iv), the addition of edges between the copy vertex and existing vertices other than the adjacent vertices of vertex $i$ is allowed, with $\beta \in [0,1]$ called the \textit{mutation rate}.

\vspace{-0.1in}
\subsubsection{Mean vertex degree.}

To obtain an analytic form for the mean vertex degree one can start by considering this model as a natural extension of the partial duplication model (Sec.~\ref{sec_partialdup}) with the addition of the mutation rate. Thus, one can write the following difference equation

\begin{equation}
	\langle E_{t+1} \rangle = \langle E_{t} \rangle + \langle k_t \rangle - \delta \langle k_t \rangle  + (t - \langle k_t \rangle) \beta.
\end{equation}

Then, as usual, by using $2 \langle E_t \rangle /t = \langle k_t\rangle$, one can recast it as

\begin{equation}
	(t+1)\langle k_{t+1} \rangle = (t) \langle k_t \rangle + 2\langle k_t \rangle - 2\delta \langle k_t \rangle + 2(t-\langle k_t \rangle) \beta,
\end{equation}

and then as

\begin{equation}
	(t+1)\langle k_{t+1} \rangle = (t+1) \langle k_t \rangle + \langle k_t \rangle - 2\delta \langle k_t \rangle + 2(t-\langle k_t \rangle) \beta.
\end{equation}

Solving this recurrence gives the following scaling with $t$

\begin{equation}
	\langle k_t \rangle \sim \frac{\beta}{\delta + \beta}t + C_{t_0}t^{1-2\delta -2\beta},
	\label{eq_beta}
\end{equation}

where $C_{t_0}$ is an integration constant. The authors of \cite{sole2002model,pastor2003evolving} proposed that one can assume $\beta = \tilde{\beta} /t$ when the mutation rate might be almost negligible compared to $\delta>1/2$. This assumption simplifies (\ref{eq_beta}) by not considering a term $\mathcal{O}(t^{-2})$ that would decay faster for large $t$, giving

\begin{equation}
	\langle k_t \rangle \sim \frac{2\tilde{\beta}}{2\delta - 1} + C_{t_0}t^{1-2\delta}.
	\label{eq_betatilde}
\end{equation}

Note that Eq.~(\ref{eq_betatilde}) returns again (\ref{sing}) when $\tilde{\beta}=0$. Since Eq.~(\ref{eq_betatilde}) has values only for $\delta \neq 1/2$, when $\delta = 1/2$ the mean vertex degree scales with $t$ as $\langle k_t \rangle \sim \tilde \beta\text{ln}(t)$.  Note that,  as pointed out in \cite{pastor2003evolving}, when a term $\mathcal{O}(t^{-2})$ appears in the rate equation, one gets

\begin{equation}
	\langle k_t \rangle \sim t^{1-2\delta}e^{2\tilde{\beta} / t} \left[ C_{t_0} + (2\tilde{\beta})^{2\delta} \Gamma(1-2\delta,2\tilde{\beta} / t) \right],
	\label{pastorwithgamma}
\end{equation}

with $C_{t_0}$ an integration constant, and $\Gamma (x,y)$ the incomplete Gamma function. In Ref. \cite{pastor2003evolving}, it is shown that for $\delta > 1/2$ and for $t \rightarrow \infty$, Eq~(\ref{pastorwithgamma}) and Eq.~(\ref{eq_betatilde}) yield the same asymptotic mean vertex degree $\langle k_{\infty} \rangle \sim 2\tilde{\beta} / (2\delta - 1)$.

\subsubsection{Mean number of edges.}

Reminiscent of (\ref{meanedge_ispo}) but with the difference of having $\nu =1$ and the additional $\beta$ term due to mutation, a continuum approximation for the evolution of the mean number of edges is written in \cite{kim2002infinite} as

\begin{equation}
	\frac{d\langle E_t \rangle}{dt} = 2(1-\delta)\frac{\langle E_t \rangle}{t} + \tilde{\beta}.
	\label{rate_edges_kim}
\end{equation}
Solving it yields the scaling with $t$

\begin{equation}
\langle
E_{t} \rangle \sim
	\begin{cases}
		t, \hspace{2.1cm}
		\mathrm{for \; } 1/2 < \delta < 1,\\
		t\tilde{\beta}\mathrm{ln}(t), \hspace{1.05cm} \mathrm{for \; } \delta = 1/2, \\
		t^{2-2\delta},  \hspace{1.45cm}\mathrm{for \; } 0<\delta<1/2.\\
	\end{cases}
	\label{edge_cases_kim}
\end{equation}

Note that the special case with $\delta = \tilde{\beta}=0$ reduces this model to a full duplication model with all the features already mentioned (Sec.~\ref{sec_fulldup}) (e.g., lack of self-averaging); also note that, when $\delta = 1$, the rate equation reads as $d\langle E_t \rangle / dt = \tilde{\beta}$, which gives a linear scaling $\langle E_t \rangle \sim  \tilde{\beta} t$.

\subsubsection{Vertex degree distribution.}

Here, the derivation of the vertex degree distribution via a rate equation is reminiscent of Eq.~(\ref{mastdeg1}) introduced in Sec.~\ref{subsec_ispo_degdist}. Yet, here, there are three modifications: (i) $\nu=1$, (ii) the contribution of the mutation rate $\beta$ and, therefore, (iii) a slightly different form of $\mathcal{M}_k$. Note that a slow mutation rate $\tilde{\beta}$ may be typically assumed, such that only a single edge adds because of mutation. Then, the rate equation for the number of vertices of degree $k$ reads as

\begin{equation}
	\frac{dN_k}{dN} = (1-\delta) \left[(k-1)\frac{N_{k-1}}{N} - k\frac{N_k}{N} \right] + \tilde{\beta} \frac{N_{k-1}}{N} - \tilde{\beta} \frac{N_{k}}{N} + \mathcal{M}_k.
	\label{rate_kim}
\end{equation}

Before expliciting the form of $\mathcal{M}_k$, one can notice that this rate equation can be conveniently written by defining $A_k = (1-\delta)k + \tilde{\beta}$:

\begin{equation}
	\frac{dN_k}{dN} = A_{k-1}\frac{N_{k-1}}{N} - A_k \frac{N_k}{N} + \mathcal{M}_k.
\end{equation}
$A_k$ is the attachment rate of a new vertex $i'$ to existing vertices. Notice that $A_k \propto k$ when $\delta \rightarrow 1^-$, therefore, linear preferential-attachment naturally emerges as a by-product of duplication (and divergence). This observation supports the hypothesis of the duplication (and divergence) principle among putative origins of linear preferential-attachment, a feature that will be deepened in the discussion section. By assuming $N_k(N) = Nn_k$, one can write

\begin{equation}
	n_k (1 + A_{k}) = A_{k-1}n_{k-1} + \mathcal{M}_k.
	\label{red_mast_kim}
\end{equation}

Concerning $\mathcal{M}_k$, this term considers a vertex of degree $k$ that joins the graph having a number $a$ of edges due to duplication, and a number $b=a-k$ edges due to mutation through two independent probabilities of realization, whose product gives (from \cite{kim2002infinite})

\begin{equation}
	\mathcal{M}_k = \sum_{a+b=k}\sum_{s=a}^{\infty} n_s \binom{s}{a} \delta^{s-a}(1-\delta)^a \frac{\tilde{\beta}^b}{b!}e^{-\tilde{\beta}}.
	\label{Gk_kim}
\end{equation}

Note that this equation simply extends (\ref{Gk_ispo}). It turns out that for $b \rightarrow 0$ (i.e., $a \approx k$), recalling that mutation edges may realistically appear at a much slower rate than the divergence rate $\delta$, Eq.~(\ref{Gk_kim}) can be approximated by (\ref{Gk_ispo}), i.e., $\mathcal{M}_k \approx (1-\delta)^{-1}n_{k/(1-\delta)}$, since the summand is sharply peaked around $s \approx k/(1-\delta)$ (from \cite{kim2002infinite}). Then, by substituting $A_k = (1-\delta)k + \tilde{\beta}$ and by assuming power-law scaling $n_k \sim k^{-\gamma}$ in Eq.~(\ref{red_mast_kim}) (as in \cite{kim2002infinite}), one can obtain a relation between the exponent $\gamma$ and the divergence rate $\delta$

\begin{equation}
	\gamma(\delta) = 1 + \frac{1}{1-\delta}-(1-\delta)^{\gamma -2}.
	\label{kim_scalw}
\end{equation}

Noteworthy, the independence of the exponent $\gamma$ from the mutation rate $\tilde{\beta}$.

\subsubsection{Vertex degree distribution with $\delta \ll 1$.}

The authors of \cite{pastor2003evolving} derived an approximation of the rate equation for the vertex degree distribution, resulting into the same form -- a power-law with exponential decay -- of an early study on the vertex degree distribution of protein-protein interaction networks \cite{jeong2001lethality}.

This derivation, however, has a number of simplifying assumptions, which are discussed in detail in the appendix of \cite{pastor2003evolving}. The rate equation is argued to be a plausible approximation only for $\delta \ll 1$ \cite{pastor2003evolving}, at which there is a chance of removing only one edge from the copy vertex due to the divergence rate $\delta$; also, the mutation process results in a simplified form \cite{pastor2003evolving}.

The rate equation for the number of vertices with degree $k$ reads as

\begin{equation}
	\frac{dN_k}{dN} = \delta \left[(k+1)\frac{N_{k+1}}{N} - k\frac{N_k}{N} \right] + (1-\delta) \left[(k-1)\frac{N_{k-1}}{N} - k\frac{N_k}{N} \right] + 2 \tilde{\beta}\left(  \frac{N_{k-1}}{N} -  \frac{N_{k}}{N}\right) + \mathcal{M}_k.
	\label{rate_pastor}
\end{equation}

One can notice the differences with Eq.~(\ref{rate_kim}). Here, $\mathcal{M}_k$ is approximated by

\begin{equation}
	\mathcal{M}_k = \frac{N_k}{N} \left(1 - k\delta \right) + \frac{N_{k+1}}{N}(k+1)\delta + \frac{N_k}{N}.
\label{gk_pastor}
\end{equation}

The first term on rhs of (\ref{gk_pastor}) is an approximation for

\begin{equation}
	 \binom{s=k}{k}\delta^{s-k}(1-\delta)^k \frac{N_s}{N} = \frac{N_k}{N}(1-\delta)^k \approx \frac{N_k}{N} \left(1 - k\delta \right),
\end{equation}

which could be plausible if one considers that only a single edge attached to the copy vertex can be lost due to divergence, plausibly occurring when $\delta \ll 1$. Then, the second term of rhs of Eq.~(\ref{gk_pastor}) is an approximation for

\begin{equation}
\binom{s=k+1}{k}\delta^{s-k}(1-\delta)^{k}\frac{N_s}{N}=\frac{N_{k+1}}{N}(k+1)\delta (1-\delta)^k \approx \frac{N_{k+1}}{N} (k+1) \delta,
\end{equation}

plausible for $\delta \ll 1$, too. The third term on rhs of (\ref{gk_pastor}) is due to unaltered duplication event contributing to increasing the number of vertices with degree $k$ by duplication of a $k$-degree vertex \cite{pastor2003evolving}. Assuming the aforementioned approximations, and considering $N_k(N) = N n_k$, one can rewrite Eq.~(\ref{rate_pastor}) in the following form

\begin{equation}
	n_{k+1}(k+1)\delta + n_{k-1}(k-1)(1-\delta) + n_{k-1}(2\tilde{\beta}) - n_k(k+2\tilde{\beta}) = 0.
	\label{rate_pastor_nk}
\end{equation}

This equation can be nicely solved with the generating function approach \cite{wilf2005generatingfunctionology}. Thus, the generating function $n_z$ is introduced as

\begin{equation}
n_z = \sum_kz^kn_k.
\end{equation}

Multiplying Eq.~(\ref{rate_pastor_nk}) for $z^k$, summing over all $k$, then from properties of generating functions one gets

\begin{equation}
	\frac{dn_z}{dz}[\delta + (1-\delta)z^2 - z] + n_z(z-1)(2\tilde{\beta}) = 0.
\end{equation}

Solving the above ordinary differential equation with boundary condition $n_{z=1}=1$, it yields

\begin{equation}
	n_z= \left[ \frac{2\delta - 1}{\delta -z(1-\delta)} \right]^{2\tilde{\beta}/(1-\delta)}.
	\label{gen_fun_pastor}
\end{equation}

As shown in \cite{pastor2003evolving}, from this expression for $n_z$, with a Taylor expansion of $n_z$ around $z=0$, one can get $n_k$ in the form of a power-law with exponential decay

\begin{equation}
	n_k = \frac{1}{k!}\frac{d^nn_z}{dz^n}\biggr\rvert_{z=0} \sim k^{-\gamma}e^{-k/k_c},
\end{equation}

which is obtained after calculations that make use of the Euler's Gamma function and Stirling approximation \cite{pastor2003evolving}. The authors of \cite{pastor2003evolving} provide the power-law exponent as

\begin{equation}
	\gamma = 1- \frac{2\tilde{\beta}}{1-\delta},
\end{equation}

and the exponential decay exponent  $k_c = \left[ \mathrm{ln}\left( \frac{\delta}{\delta -1}\right) \right]^{-1}$. When $\delta \ll 1$ assumption does not hold, the aforementioned approximations may not be valid, as shown in the appendix of Ref.~\cite{pastor2003evolving}.

\subsection{Duplication-divergence with mutation and dimerization}
\label{DDHM}

\begin{figure}[t!] \vspace{0.2in}
	\hspace{2.5cm}
	\includegraphics[width=0.73\textwidth]{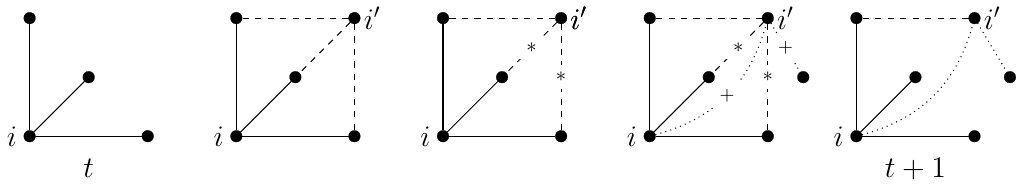}
	\caption{Simplified depiction of a possible realization of an iteration of the growing graph model by duplication-divergence with mutation and dimerization. Dashed edges attached to $i'$ have been duplicated from vertex $i$. Edges marked with $*$ are not conserved at $t+1$ due to the divergence process. Dotted edges marked with $+$ are added due to mutation and dimerization.}
	\label{dd_as_h_m_fig}
\end{figure}

The growing graph model here reviewed combines duplication and complete asymmetric divergence with two additional sophistications, namely, \textit{mutation} (introduced in Sec.~\ref{sec_dupdivmut}) and \textit{dimerization}. Dimerization takes the name from molecular biology, when two proteins self-associate to form a dimer. In graph theoretic terms, it is abstracted by the addition of an edge between vertex $i$ and vertex $i'$. The generic iteration for the duplication and divergence model with both dimerization and mutation is (see also Fig.~\ref{dd_as_h_m_fig}):

\begin{tcolorbox}[boxrule=0.8pt, colframe=white, colback=white, sharp corners]
\begin{enumerate}[leftmargin=30pt,itemsep=-0.01in]
	\item Choice of a vertex $i$ uniformly at random among existing vertices.
	\item Duplication of vertex $i$ into a new vertex $i'$, having the same edges of $i$.
	\item Each duplicate edge $(i',j)$ is conserved with probability $p$ (namely, lost with probability $1-p=\delta$).
	\item An edge between vertex $i'$ and any other vertex of the graph (except $i$'s adjacent vertices) is added with probability $\beta$.
	\item  Edge $(i,i')$ is added with probability $\alpha$.
\end{enumerate}
\end{tcolorbox}

Here, (v) allows the addition of the dimerization link between the original vertex and the copy vertex; $\alpha \in [0,1]$ is the \textit{dimerization rate}. This model includes two sources of addition of new edges (dimerization and mutation) other than those that are duplicated \cite{cai2015mean}.

\subsubsection{Mean vertex degree.}
To obtain the mean vertex degree, one can begin by writing the following equation

\begin{equation}
	\langle E_{t+1} \rangle =  \langle E_{t} \rangle + \langle k_t \rangle - \delta \langle k_t \rangle + \alpha + \beta(N-\langle k_t \rangle -1)
	,
\end{equation}	

The authors of \cite{cai2015mean} assumed a slow mutation rate $\tilde{\beta}$ (a single edge added)

\begin{equation}
	\langle E_{t+1} \rangle =  \langle E_{t} \rangle + \langle k_t \rangle - \delta \langle k_t \rangle + \alpha + \tilde{\beta}.
	\label{edgs_mut_dim}
\end{equation}

Then, as typically carried out,  $\langle k_t \rangle=2 \langle E_t \rangle/ t$ is used to recast the above equation as follows

\begin{equation}
	(t+1)\langle k_t \rangle = (t+1)\langle k_t \rangle + \langle k_t \rangle - 2\delta \langle k_t \rangle + 2(\alpha + \tilde{\beta}).
\end{equation}

Solving the recurrence yields the scaling

\begin{equation}
	\langle k_t \rangle \sim \frac{2\alpha}{2\delta -1 }+\frac{2\tilde{\beta}}{2\delta -1}+C_{t_0}t^{1-2\delta},
	\label{avg_cai}
\end{equation}

which has value for $\delta \neq 1/2$. For $\delta =1/2$ the mean vertex degree has the form $\langle k_t \rangle \sim (\alpha + \tilde{\beta})\mathrm{ln}(t)$. Note that when $\alpha=0$ (absence of dimerization), (\ref{avg_cai}) becomes Eq.~(\ref{eq_betatilde}).

\subsubsection{Mean number of edges.}

The mean number of edges follows directly from (\ref{edgs_mut_dim})

\begin{equation}
	\langle E_{t+1} \rangle =  \langle E_{t} \rangle + \langle k_t \rangle - \delta \langle k_t \rangle + \alpha + \tilde{\beta},
\end{equation}

that can be written as

\begin{equation}
	\langle E_{t+1} \rangle - \langle E_{t} \rangle = 2(1-\delta)\frac{\langle E_t \rangle}{t} + \alpha + \tilde{\beta}.
\label{cai_edges}
\end{equation}

Note that with a continuum approximation $d\langle E_t \rangle/dt \simeq \langle E_{t+1} \rangle - \langle E_{t} \rangle$, (\ref{cai_edges}) is reminiscent of Eq.~(\ref{rate_edges_kim}) with an additional term $\alpha$ due to dimerization. Solving it gives

\begin{equation}
	\langle E_t \rangle \sim \frac{\alpha}{2\delta -1}t + \frac{\tilde{\beta}}{2\delta - 1}t+C_{t_0}t^{2-2\delta},
\end{equation}

with $C_{t_0}$ an integration constant. The above solution holds for $\delta \neq \frac{1}{2}$. Then, for $\delta = \frac{1}{2}$, one gets $\langle E_t \rangle \sim t(\alpha + \tilde{\beta})\mathrm{ln}(t) + C_{t_0}t$. It is worth noting that, when $\alpha=0$, results shown in (\ref{edge_cases_kim}) are returned.

\subsubsection{Vertex degree distribution.}

The vertex degree distribution $n_k$ can be derived by starting with a rate equation (see, \cite{cai2015mean}) describing the evolution of the number of vertices with degree $k$, $N_k(N)$

\begin{equation}
\begin{aligned}
	\frac{dN_k}{dN} &= \left[ \alpha + (1-\delta)(k-1) + \tilde{\beta}\left(1 - \frac{kN_{k-1}}{N} \right)  \right] \frac{N_{k-1}}{N} - \\ & - \left[ \alpha + (1-\delta)k + \tilde{\beta}\left(1 - \frac{(k+1)N_{k}}{N} \right)  \right]\frac{N_k}{N} + \mathcal{M}_k,
\end{aligned}
\label{mast_cai}
\end{equation}

with $\mathcal{M}_k$ given by

\begin{equation}
\begin{aligned}
\mathcal{M}_k = &(1-\alpha)\sum_{m=0}^{\infty}\sum_{s=k-m}^{\infty} \frac{N_s}{N} \binom{s}{k-m}(1-\delta)^{s-(k-m)} \delta^{k-m} + \\  &+ \alpha  \sum_{m=0}^{\infty}\sum_{s=k-1-m}^{\infty} \frac{N_s}{N} \binom{s}{k-1-m}(1-\delta)^{s-(k-1-m)} \delta^{k-1-m}.
\end{aligned}
\label{G_kcai}
\end{equation}

If one assumes convergence for large $N$, then $dN_k/dN = n_k$ as $N_k(N) =N n_k$. Then, as in \cite{cai2015mean}, one can approximate the rate equation by neglecting quadratic terms (as they decay faster for large $N$), and write Eq.~(\ref{mast_cai}) as

\begin{equation}
	\left[ \alpha + (1-\delta)(k-1) + \tilde{\beta}  \right] \frac{n_{k-1}}{n_k} - \left[ \alpha + (1-\delta)k + \tilde{\beta} \right] + \mathcal{M}_k / n_k =0,
\end{equation}

Assuming a power-law scaling $n_k \sim k^{-\gamma}$, one obtains

\begin{equation}
	\left[ \alpha + (1-\delta)(k-1) + \tilde{\beta}  \right] \left( \frac{k}{k-1} \right)^\gamma - \left[ \alpha + (1-\delta)k + \tilde{\beta} \right] + \tilde{\mathcal{M}}_k=0,
\end{equation}

with $\tilde{\mathcal{M}}_k = \mathcal{M}_k / k^{-\gamma}$. Considering results of Ref.~\cite{bebek2006degree} one can recast the first two terms as $-1+(\gamma-1)(1-\delta) + \mathcal{O}(k^{-1})$, and $\tilde{\mathcal{M}}_k$ as $[1+\mathcal{O}(k^{-1)}](1-\delta)^{\gamma -1}$ (see, Ref.~\cite{cai2015mean}). Finally, for high $k$ values (tail of $n_k$) one obtains \cite{cai2015mean}

\begin{equation}
\gamma(\delta) = 1 + \frac{1}{1-\delta}-(1-\delta)^{\gamma -2}.
\end{equation}

This relation recalls  (\ref{kim_scalw}) obtained by the authors of \cite{kim2002infinite} (that, intriguingly, does not directly include dimerization). Note, indeed, the independence from $\alpha$ and $\tilde{\beta}$.

\subsubsection{Mean-field considerations, clustering, (dis)assortativity.}

The model introduced at the beginning of Sec.~\ref{DDHM} considers only the special case of complete asymmetric divergence, yet, through a mean-field approach, the model seems to well synthesize other duplication (and divergence) models. Indeed, in many cases, within the mean-field approximation, complete asymmetric divergence and symmetric divergence collapse onto the same rate equation. For instance, consider the evolution of $\langle E_t \rangle$ for a simple partial duplication with symmetric divergence (with non-interacting vertices). Recall that symmetric divergence is a divergence process where each duplicate edge can be lost either by the copy vertex or by the original vertex, with same probability. Therefore, if $j$ is an adjacent vertex of $i$, then after duplication, only one of the two edges $(i,j)$, $(i',j)$ is lost with probability $\delta$.  The total probability of removing $(i,j)$ is $\delta/2$, and that of removing $(i',j)$ is $\delta/2$ as well, as one choses \textit{where} to remove, which is from $i$ (or from $i'$) with probability $1/2$, and \textit{if} to remove the edge with probability $\delta$, combining as $\delta/2$. Then, the rate equation for the mean number of edges reads as

\begin{equation}
	\langle E_{t+1} \rangle = \langle E_t \rangle  + \langle k_t \rangle - \frac{\delta}{2}\langle k_t \rangle - \frac{\delta}{2}\langle k_t \rangle=\langle E_t \rangle  + \langle k_t \rangle -\delta\langle k_t \rangle,
\end{equation}

which returns Eq.~(\ref{partdup_begin}). Hence, the mean-field rate equation for $\langle E_t \rangle$ in the partial duplication model with symmetric divergence returns the same mean-field rate equation that we have seen in the partial duplication model with complete asymmetric divergence (with non-interacting vertices). Yet, it turns out that symmetric divergence might microscopically have different behavior from complete asymmetric divergence, yet being macroscopically similar.

One can explore additional structural characteristics through mean-field rate equations concerning the models that have been introduced so far. For the duplication-divergence-dimerization-mutation model (Sec.~\ref{DDHM}), one of these structural characteristics is the mean clustering coefficient $\langle C(k_i) \rangle$ of a vertex $i$ (chosen uniformly at random among vertices of the graph), which gives the mean probability that two adjacent vertices of vertex $i$ are themselves adjacent.

An analytic form of the mean clustering coefficient of the duplication-divergence-dimerization-mutation model is shown by the authors of \cite{cai2015mean}. They begin by considering $\langle k_a \rangle$, i.e. the vertex degree of a random uniform vertex $a$, and writing the following rate equation

\begin{equation}
	\frac{d\langle k_a \rangle}{dt} = \frac{\alpha}{t} + (1-\delta)\frac{\langle k_a \rangle}{t} + \left( 1 - \frac{\langle k_a \rangle + 1}{t} \right)\frac{\tilde{\beta}}{t}.
\end{equation}

The first term is due to an increase in $\langle k_a \rangle$ as a vertex $a$ is duplicated into $a'$ and one edge adds because of dimerization. The second term accounts for the case when one of the adjacent vertices of $a$, say $b$, is duplicated into $b'$ (which happens with probability $\langle k_a \rangle/t$), and $(a,b')$ is kept with probability $(1-\delta)$. The third term  adds to $\langle k_a \rangle$ when a vertex other than vertex $a$ and one of its $\langle k_a \rangle$ adjacent vertices is duplicated, and an edge is added by mutation with vertex $a$ as one of the two edge ends. The authors of \cite{cai2015mean} consider the following approximation that neglects $\mathcal{O}(t^{-2})$ terms, yielding

\begin{equation}
	\frac{d\langle k_a \rangle }{dt} \simeq \frac{(1-\delta)}{t}\left[\langle k_a \rangle  + \frac{\alpha + \tilde{\beta}}{(1-\delta)} \right] = \frac{(1-\delta)}{t}(\langle k_a \rangle +\eta),
	\label{ka_rate}
\end{equation}

denoting $\eta = (\alpha + \tilde{\beta})/(1-\delta)$.  Then, a rate equation of the total number of edges between the adjacent vertices of $a$, denoted by $\langle g_a \rangle$ is written as in \cite{cai2015mean}

\begin{equation}
	\frac{d\langle g_a \rangle}{dt} = (1-\delta) \left\{ \frac{\langle k_a \rangle}{t} \alpha + \frac{\langle k_a \rangle}{t} \left[ \alpha +C(k_a)(\langle k_a \rangle -1)(1-\delta) + \mathcal{O}(t^{-2})  \right] \right\}.
\end{equation}

The first term in brackets can be explained by considering that vertex $a$ is duplicated into vertex $a'$ with dimerization occurring, thus forming a triangle with vertex $b$ (adjacent vertex of $a$), whose edge $(b,a')$ is conserved with probability $(1-\delta)$, hence giving $(1-\delta)\alpha$ multiplied by the fraction of adjacent vertices of $a$ (i.e., $\langle k_a \rangle/t$).

The second term in brackets is non-trivial: one needs to consider the duplication of a vertex $b$ adjacent to vertex $a$, duplicated into $b'$, whose edge $(b',a)$ is conserved with probability ($1-\delta$), and also the dimerization edge $(b,b')$ is added to form a triangle with $a$ resulting in the first term of the second term in brackets (i.e., $\alpha(1-\delta) \langle k_a \rangle/t$). Then, the non-trivial part here is that one has to consider the case in which an adjacent vertex $j'$ of both $a$ and $b$, which is thus adjacent of $a$ and duplicated into $b'$, can form the triangle ($a, j',b'$) when the edge $(b',j')$ is kept with probability $(1-\delta)$. The edge $(b',a)$ is kept with probability $(1-\delta)$ as well, hence the probability of this event occurring is $(1-\delta)^2$ multiplied by the probability of having two adjacent vertices of $a$ connected (i.e., $C(k_a)$) times all possible $j'$ vertices, overall occurring with probability $(1-\delta)^2C(k_a)(\langle k_a \rangle -1)$. The last term is an $\mathcal{O}(t^{-2})$ term that, in \cite{cai2015mean}, it was neglected for large $t$, giving

\begin{equation}
	\frac{d\langle g_a \rangle}{dt} \simeq \frac{(1-\delta)}{t} \left\{ \langle k_a \rangle \alpha + \langle k_a \rangle \left[ \alpha +C(k_a)(\langle k_a \rangle -1)(1-\delta)  \right] \right\}.
\end{equation}

Using chain rule, recalling Eq.~(\ref{ka_rate}), and replacing $C(k_a) = 2\langle g_a \rangle /\langle k_a \rangle (\langle k_a \rangle - 1)$, one gets \cite{cai2015mean}

\begin{equation}
	\frac{d\langle g_a \rangle}{d\langle k_a \rangle} = \frac{d\langle g_a \rangle}{dt} \frac{dt}{d\langle k_a \rangle} \simeq \frac{2(1-\delta)}{k_a + \eta} \langle g_a \rangle - \frac{2\langle k_a \rangle \alpha}{\langle k_a \rangle  + \eta}.
\end{equation}

Solving this rate equation for $\langle g_a \rangle $, and substituting it into the $g$ appearing in the definition of clustering coefficient, gives a power-law scaling \cite{cai2015mean}

\begin{equation}
	\langle C(k) \rangle  = \left\langle \frac{2g}{k(k-1)} \right\rangle \sim k^{-\lambda_C},
\end{equation}

which can be specialized to different values of $\lambda_C$ depending on $\alpha$ and $\delta$ (see Ref. \cite{cai2015mean}).

Similarly, a rate equation can be obtained for the average number of edges among adjacent vertices of vertex $a$ \cite{cai2015mean}, i.e., $\langle k_{nn}(a) \rangle$: the authors of \cite{cai2015mean} suggests a power-law scaling $\langle k_{nn}(a) \rangle \sim k^{-\lambda_D}$, with $\lambda_{D} \geq 0$ non-negative, which may not suggest assortative mixing for graphs generated by this model \cite{cai2015mean}.

\subsection{Duplication-divergence (symmetric coupled) with dimerization}
\label{sec_dupdivdim_coup}
	
\begin{figure} \vspace{0.2in}
	\hspace{2.5cm}
	\includegraphics[width=0.72\textwidth]{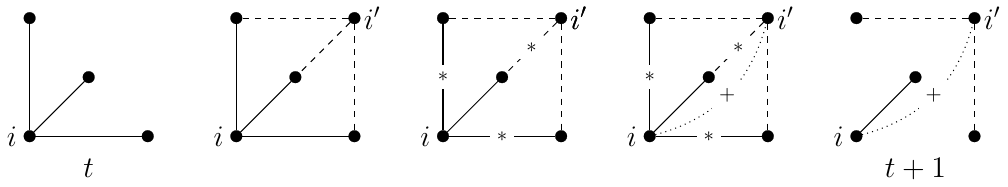}
	\caption{Simplified depiction of a possible realization of an iteration of the duplication, (symmetric coupled) divergence with dimerization model. Dashed edges attached to $i'$ have been duplicated from vertex $i$. Edges marked with $*$ are not conserved due to the divergence process. Dotted edges marked with $+$ are added due to mutation and dimerization. Note that, due to the coupled symmetric divergence, both $i$ and $i'$ can lose edges due to the divergence process.}
	\label{dd_sc_h_fig}
\end{figure}

This model encompasses a divergence process that is symmetric with respect to duplicate vertices, meaning that duplicate edges can be lost both from the original vertex $i$ and from the copy vertex $i'$; also, the symmetric divergence process is \textit{coupled}, meaning that each duplicate edge can be lost either from $i$ or from $i'$, but not from both \cite{vazquez2003modeling}. The generic iteration for this model is the following (see also Fig.~\ref{dd_sc_h_fig}):

\begin{tcolorbox}[boxrule=0.8pt, colframe=white, colback=white, sharp corners]
\begin{enumerate}[leftmargin=30pt,itemsep=-0.01in]
	\item Choice of a vertex $i$ uniformly at random among existing vertices.
	\item Duplication of vertex $i$ into vertex $i'$, having the same edges of $i$.
	\item For each couple of duplicate edges $\lbrace (i,j),(i',j) \rbrace$, one of the two edges is randomly chosen with probability $1/2$ and then it is lost with probability $1-p=\delta$.
	\item Edge ($i,i'$) is added with probability $\alpha$.
\end{enumerate}
\end{tcolorbox}

Note that, even when $\delta=1$, the divergence process described above in (iii) allows to overall conserve original edges (namely, edges of $i$), complementarily distributing them among the copy vertex and the original vertex. 

Concerning the vertex degree distribution for this model, no prior studies actually provide an analytic vertex degree distribution. Yet, as shown in \cite{vazquez2003growing}, moments of the time-dependent vertex degree distribution have shown to exhibit \textit{multifractality}, meaning that the vertex degree distribution cannot be characterized by a single scaling exponent. Multifractality, a peculiar feature of this model, will be further detailed in the discussion section, as it shows a  very peculiar property of duplication (and divergence) growing graph models.

\subsubsection{Mean vertex degree.}

To get the mean vertex degree, one can follow the same procedure described in Sec.~\ref{sec_partialdup}, with an additional term that
	concerns the dimerization rate $\alpha$. Indeed, one can begin by writing the following
	
\begin{equation}
	\langle E_{t+1} \rangle =  \langle E_{t} \rangle + \langle k_t \rangle - \delta \langle k_t \rangle + \alpha,
	\label{heter_begin}
\end{equation}			
	
whose first three terms on rhs are the same as those in Eq.~(\ref{partdup_begin}), with an additional $\alpha$ term accounting for the addition of one link $(i,i')$ with probability $\alpha$. It is important to note that despite having a divergence process that is symmetric coupled (i.e., a loss of duplicate edge from $i$ is a non-overlapping event from a loss of duplicate edge from $i'$), in the mean-field description, the loss term is still $-\delta \langle k_t \rangle$, the same as in the asymmetric divergence process. Indeed, for a duplicate edge pair $(\fullin)$ one can consider the probabilities of transitioning to the configuration indicated in parentheses on lhs:

\begin{equation}
\begin{aligned}
  P( \fullin ) &= \frac{1}{2}(1-\delta) + \frac{1}{2}(1-\delta) = 1-\delta  \\
	 P( \copiedout ) &= \delta /2  \\
	 P( \origout )
	 &= \delta/2,
\end{aligned}
\label{probs_coup}
\end{equation}

and notice that, in coupled symmetric divergence, the graph loses on average each of $\langle k_t \rangle$ from vertex $i$ with probability $\delta /2$, and each $\langle k_t \rangle$ from vertex $i'$ with probability $\delta/2$ as well. In terms of the average loss of edges at iteration $t$, one gets

\begin{equation}
	\langle k_t \rangle[P( \copiedout ) + P( \origout ) ] = \langle k_t \rangle[\delta/2 + \delta/2] = \delta\langle k_t \rangle,
\end{equation}

which explains the third term in the rhs of Eq.~(\ref{heter_begin}), i.e., $-\delta\langle k_t \rangle$. As usual, we can use $\langle k_t \rangle = 2\langle E_t \rangle /t$ and write Eq.~(\ref{heter_begin}) as

\begin{equation}
	(t+1)\langle k_{t+1} \rangle  = (t+1)\langle k_t \rangle + \langle k_t \rangle -2\delta\langle k_t \rangle + 2\alpha.
\end{equation}

Solving the above recurrence one gets the following \cite{vazquez2003modeling,sole2008spontaneous}

\begin{equation}
	\langle k_t \rangle \sim \frac{2\alpha}{2\delta-1} + C_{t_0}t^{1-2\delta},
	\label{sols_sub}
\end{equation}

with $C_{t_0}$ an integration constant that depends on the vertex degree of the initial graph. For $\alpha =0$ (no dimerization), one can recover the partial duplication solution, i.e., (\ref{sing}), which is intriguing as it shows that a mean-field description yields the same form of the average vertex degree in both the partial duplication with complete asymmetric divergence and in symmetric coupled divergence with no dimerization.

\subsubsection{Mean number of edges.}

To obtain the mean number of edges, one can start from Eq.~(\ref{heter_begin}), and substitute $\langle k_t \rangle = 2\langle E_t \rangle/t$ to obtain

\begin{equation}
	\langle E_{t+1} \rangle =  \langle E_{t} \rangle + 2 \frac{\langle E_t \rangle}{t} - \delta \frac{\langle E_t \rangle}{t} + \alpha.
\end{equation}

which, with a continuum approximation, one can write this recurrence as

\begin{equation}
	\frac{d\langle E_t \rangle}{dt} = 2(1-\delta)\frac{\langle E_t \rangle}{t} + \alpha.
\end{equation}

The rate equation obtained here for the mean number of edges is the same form of Eq.~(\ref{rate_edges_kim}), with same solution form too, yet, here, the last term of the right end side of the rate equation is the dimerization rate $\alpha$ instead of the mutation rate $\beta$. Solving the rate equation, one obtains the following scaling with $t$

\begin{equation}
\langle E_{t} \rangle \sim \frac{\alpha t}{2\delta-1} + C_{t_0} t^{2 - 2 \delta},
\label{sols}
\end{equation}

with $C_{t_0}$ the integration constant that depends on initial conditions. Note that another way to obtain (\ref{sols}) is by substituting $\langle E_t \rangle = t \langle k_t \rangle / 2$ into (\ref{sols_sub}).

\subsubsection{Vertex degree distribution.}
\label{Sec_multifract}

As anticipated, in \cite{vazquez2003growing}, the vertex degree distribution of the duplication model with coupled divergence and dimerization was not directly solved but a general master equation is written, and moments of the distribution can be calculated. The rate equation is

\begin{equation}
	t\frac{dn_k}{dt} + n_k = (1-\delta) \left[(k-1)n_{k-1} - k n_k  \right] + \alpha \left[ n_{k-1} - n_k \right] + 2\alpha \mathcal{M}_{k-1} +2(1-\alpha)\mathcal{M}_{k}.
	\label{mast_vazq}
\end{equation}

Here, $\mathcal{M}_k$ assumes the following form

\begin{equation}
	\mathcal{M}_k = \sum_{s>k} \binom{s}{k} \left(\frac{\delta}{2}\right)^{s-k} \left( 1-\frac{\delta}{2}\right)^{k} n_s.
\end{equation}

From Eq.~(\ref{mast_vazq}), one can calculate $l$th-moments of the distribution $n_k$ by summing over $k$ and by multiplying by $k^l$ both rhs and lhs. The moments equation reads as \cite{vazquez2003growing}

\begin{equation}
	\langle k^l \rangle = \sum_kk^ln_k \propto t^{\tau(\delta,l)},
	\label{moments_eq}
\end{equation}

with the exponent given by the following relation

\begin{equation}
	\tau(\delta,l) = l(1-\delta) + 2 \left[ \left(\frac{2-\delta}{2} \right)^l - 1\right].
\label{eq_MF1}
\end{equation}

The nonlinearity of the exponent $\tau(\delta,l)$ suggests a \textit{multifractal} vertex degree distribution \cite{vazquez2003growing}. Note that, with $l=1$, one can obtain the scaling of the mean vertex degree as $\tau(\delta,l=1) = 1-2\delta$ and, thus, $\langle k \rangle \sim t^{\tau(\delta,1)} \sim t^{1-2\delta} $, which is reminiscent of Eq.~(\ref{sols_sub}).

\subsection{Duplication-divergence (symmetric uncoupled) with dimerization}
\label{sec_dupdivdim_uncoup}
This model introduces the \textit{uncoupled symmetric divergence} process, meaning that a duplicate edge can be lost from both the original vertex $i$ and the copy vertex $i'$ \cite{ispolatov2005cliques,sudbrack2018master}. The generic iteration for this model is

\begin{tcolorbox}[boxrule=0.8pt, colframe=white, colback=white, sharp corners]
\begin{enumerate}[leftmargin=30pt,itemsep=-0.01in]
	\item Choice of a vertex $i$ uniformly at random among existing vertices.
	\item Duplication of vertex $i$ into vertex $i'$, having the same edges of vertex $i$.
	\item Each duplicate edge $(i',j)$ is lost with probability $\delta_{i'}$, and each original edge $(i,j)$ is lost with probability $\delta_i$.
	\item Edge ($i',i$) is added with probability $\alpha$.
\end{enumerate}
\end{tcolorbox}

In the following, the mean vertex degree, number of edges, and vertex degree distribution are recovered following the work of the authors of \cite{ispolatov2005cliques} and \cite{sudbrack2018master}. Note that, in \cite{sudbrack2018master}, the dimerization rate is always set equal to $\alpha=1$, while here $\alpha$ is arbitrary in order to generalize results of both \cite{ispolatov2005cliques} and \cite{sudbrack2018master}. Since the authors of \cite{ispolatov2005cliques} do not include non-interacting vertices in the growing graph when loss of all edges happens due to divergence, here, the mean number of edges with and without non-interacting vertices are reviewed in separately.

\subsubsection{Mean vertex degree.}

Here, it is convenient to introduce, for a duplicate edge pair $(\fullin)$, the following probabilities of transitioning to the configuration indicated within parentheses on lhs:

\begin{equation}
\begin{aligned}
	 P( \fullin ) &= (1-\delta_i)(1-\delta_{i'})  \\
	 P( \copiedout ) &= (1-\delta_i)\delta_{i'}  \\
	 P( \origout )
	 &= \delta_i(1-\delta_{i'}) \\
	 P( \bothout )
	 &= \delta_i\delta_{i'} .
\end{aligned}
\label{probs_uncoup}
\end{equation}

For the uncoupled symmetric divergence, in a generic iteration, the graph loses on average each of the $\langle k_t \rangle$ edges from vertex $i'$ each with probability $P( \copiedout )$, each of the $\langle k_t \rangle$ edges from vertex $i$ with probability $P( \origout )$, and also, the graph loses each of the $2\langle k_t \rangle$ edges from both $i$ and $i'$ with probability $P( \bothout )$, see Eqs.~(\ref{probs_uncoup}).

Then, the mean loss of number of edges at iteration $t$ is thus

\begin{equation}
	\langle k_t \rangle \left[P( \copiedout ) + P( \origout ) + P( \bothout )\right],
	\label{probab_uncop}
\end{equation}

Using Eqs.~(\ref{probs_uncoup}) in Eq.~(\ref{probab_uncop}) gives

\begin{equation}
	 \langle k_t \rangle \left[ (1-\delta_i)\delta_{i'}+\delta_i(1-\delta_{i'}) + \delta_i \delta_{i'}\right] = \langle k_t \rangle (\delta_i + \delta_{i'}).
\end{equation}

Denoting $\delta_i + \delta_{i'} = M$ as in \cite{sudbrack2018master}, one can write the rate equation for the mean vertex degree as we usually do, starting from the mean variation of the number of edges

\begin{equation}
	\langle E_{t+1} \rangle - \langle E_t \rangle = \langle k_t \rangle - M\langle k_t \rangle + \alpha.
\end{equation}

In \cite{sudbrack2018master}, $\alpha$ is set to 1. Then, we use $2\langle E_t \rangle /t = \langle k_t \rangle$, to write

\begin{equation}
(t+1) \langle k_{t+1} \rangle = (t) \langle k_t \rangle +2\langle k_t \rangle -2M\langle k_t \rangle + 2\alpha.
\label{sudbrac_red}
\end{equation}

Solving the above recurrence, one obtains

\begin{equation}
	\langle k_t \rangle \sim  \frac{2\alpha}{2M-1} + C_{t_0}t^{1-2M}.
	\label{sud_meank}
\end{equation}

Note that $M \in [0,2]$, and that this solution is restricted to $M\neq 1/2$. For $M=1/2$, the mean vertex degree scales logarithmically with $t$, namely,	$\langle k_t \rangle \sim 2\mathrm{ln}(t)$ \cite{sudbrack2018master}.

\subsubsection{Mean number of edges.}

Similarly, one can start from the rate equation of the mean number of edges \cite{sudbrack2018master}

\begin{equation}
	\langle E_{t+1} \rangle = \langle E_t \rangle + \langle k_t \rangle - M\langle k_t \rangle + \alpha,
\end{equation}

and, then, obtain the scaling with $t$ for the mean number of edges

\begin{equation}
	\langle E_t \rangle \sim \frac{\alpha t}{2M-1} + C_{t_0}t^{2-2M},
	\label{sud_et}
\end{equation}

which is valid for $M\neq 1/2$. For the case of $M=1/2$, the mean number of edges scales as $\langle E_t \rangle \sim t\mathrm{ln}(t) + C_{t_0}t$. Note that, in \cite{sudbrack2018master}, the solution appears explicitly with $(t+1)$, while here, instead, $t \gg 1$ was assumed, and thus $t+1 \simeq t$ (the reader can refer to \cite{sudbrack2018master} and compare it with (\ref{sud_et}), and also with (\ref{sud_meank})).

\subsubsection{Mean vertex degree (without non-interacting vertices).}

In \cite{ispolatov2005cliques}, the mean vertex degree is calculated by considering the rate of addition of new connected vertices, say $\nu$. In this sense, the mean vertex degree is derived by starting from a continuum form of the mean variation of the number of edges

\begin{equation}
	\nu\frac{d\langle E_N \rangle }{dN} = \frac{2\langle E_N \rangle}{N} \left( 1 - \delta_{i'}  - \delta_{i} \right) + \alpha.
	\label{nosiol_cliques}
\end{equation}

Note that $N$ is the number of vertices with at least one edge, and $t$ does not appear because the identity $t=N$ is violated due to the exclusion of non-interacting vertices after a full divergence, which occurs when all edges of a vertex (vertex $i$, vertex $i'$, or an adjacent vertex of them) are lost. Knowing that $\langle k_N \rangle = 2\langle E_N \rangle/N$, substituting it into Eq.~(\ref{nosiol_cliques}) and solving, it gives

\begin{equation}
\langle k_N \rangle \sim \frac{2\alpha}{\nu - 2(1- \delta_{i}-\delta_{i'})} + C_{t_0}N^{\frac{2}{\nu}(1-\delta_i-\delta_{i'})-1},
\label{nu_ispo}
\end{equation}
valid for $\nu \neq 2(1- \delta_{i}-\delta_{i'})$, with $C_{t_0}$ an integration constant that depends on the initial graph. One can immediately check that, for $\nu=1$, Eq.~(\ref{nu_ispo}) returns Eq~(\ref{sud_meank}). Then, for $\nu = 2(1- \delta_{i}-\delta_{i'})$. the mean vertex degree scales with $N$ as

\begin{equation}
	\langle k_N \rangle \sim 2 + \frac{\alpha \mathrm{ln}(N)}{2(1-\delta_{i} - \delta_{i'})}.
\end{equation}

obtained by substituting $\nu = 2(1- \delta_{i}-\delta_{i'})$ into Eq.~(\ref{nosiol_cliques}) and solving it.

\subsubsection{Mean number of edges (without non-interacting vertices).}

Here, the case of exclusion of non-interacting vertices from the calculation of the mean number of edges is considered. By solving Eq.~(\ref{nosiol_cliques}), one can immediately get the mean number of edges without non-interacting vertices

\begin{equation}
	\langle E_N \rangle \sim \frac{\alpha N}{\nu - 2(1- \delta_{i}-\delta_{i'})} + C_{t_0}N^{\frac{2}{\nu}(1-\delta_i-\delta_{i'})}.
	\label{cliques_E}
\end{equation}

Note that the above solution holds only for $\nu \neq 2(1- \delta_{i}-\delta_{i'})$. Also note that, when $\nu=1$ (thus, $N=t$), Eq.~(\ref{cliques_E}) becomes Eq.~(\ref{sud_et}). Then, the mean number of edges scales with $t$ as

\begin{equation}
	\langle E_N \rangle \sim \frac{\alpha N\mathrm{ln}(N)}{2(1-\delta_{i} - \delta_{i'})} + C_{t_0}N.
\end{equation}

for $\nu = 2(1- \delta_{i}-\delta_{i'})$, with $C_{t_0}$ an integration constant.  The above equation for the mean number of edges holds for $\delta_i +\delta_{i'}\neq1$, and it tells that for $\delta_i +\delta_{i'}<1$ the number of edges is vanishing when $N$ increases, as the graph loses edges faster than it gains new edges \cite{ispolatov2005cliques}.

\subsubsection{Vertex degree distribution.}

As it has been similarly shown in Sec.~\ref{subsec_ispo_degdist}, in \cite{ispolatov2005cliques}, a rate equation for the mean number of vertices with degree $k$ (i.e., $N_k$) can be written as

\begin{equation}
	\nu\frac{dN_k}{dN} = \left( 1- \delta_i - \delta_{i'} \right) \left[(k-1)\frac{N_{k-1}}{N} - k\frac{N_k}{N} \right] + \mathcal{M}_k,
	\label{remi_cliq}
\end{equation}

with $\mathcal{M}_k$ approximated as in \cite{ispolatov2005cliques,ispolatov2005duplication}, i.e.,

\begin{equation}
	\mathcal{M}_k \approx \frac{N_{k/p_i}}{N(1-\delta_i)} + \frac{N_{k/p_{i'}}}{N(1-\delta_{i'})} - \frac{N_k}{N},
\end{equation}

setting $p_i = 1-\delta_i$ and $p_{i'}=1-\delta_{i'}$ to simplify notation. Eq.~(\ref{remi_cliq}) is reminiscent of Eq.~(\ref{remi}) with proper modifications due to symmetric uncoupled divergence. Then, assuming a power-law scaling $n_k \sim k^{-\gamma}$, the authors of \cite{ispolatov2005cliques} wrote the following relation between the exponent $\gamma$ and parameters of the model

\begin{equation}
	p_i^{\gamma -1} + p_{i'}^{\gamma-1} + (p_i + p_{i'}-1)(\gamma-1)+1-2(p_i+p_{i'})=0.
\end{equation}

with a trivial solution $\gamma=2$, and a non-trivial one that depends on $p_i,p_{i'}$ as pointed out in Ref.~\cite{ispolatov2005cliques}. 

\vspace{-0.05in}
\subsection{Deletion, duplication-divergence and dimerization}
\label{sec_model_farid}

Some authors have included a \textit{deletion} process within models that were reviewed so far. This sophistication embeds a different approach (based on two time-steps) to write rate equations for the evolution of the observables of graphs growing through these network growth models. The model, introduced in \cite{farid2006evolving}, includes \textit{deletion} through a removal of a vertex (and of its edges) chosen uniformly at random among existing vertices.
The generic iteration of this model comprises two time-steps ($t \rightarrow t+2$), in the following way:
\begin{tcolorbox}[boxrule=0.8pt, colframe=white, colback=white, sharp corners]
\begin{enumerate}[leftmargin=30pt,itemsep=-0.01in]
	\item  From $t \rightarrow t+1$, a vertex $i$ chosen uniformly at random is removed (with all its edges) with probability $p_{\mathrm{del}}$.
	\item Then, from $t+1 \rightarrow t+2$, with probability $p_{\mathrm{dup}}$, the iteration of this model continues as a partial duplication with dimerization model (below: (iii) to (vi)).
	\item Choice of a vertex $i$ uniformly at random among existing vertices.
	\item Duplication of vertex $i$ into vertex $i'$, having the same edges of vertex $i$.
	\item Each duplicate edge $(i',j)$ is conserved with probability $p$ and lost with probability $1-p=\delta$.
	\item An edge $(i,i')$ is added with probability $\alpha$.
\end{enumerate}
\end{tcolorbox}	
	
When $p_{\mathrm{del}}=0$, the model is a duplication-divergence (with complete asymmetric divergence) and dimerization, which is the same model introduced in Sec.~\ref{sec_partialdup}. The novelty here is that when $p_{\mathrm{del}}\neq0$, evolution equations can be written as two-step rate equations as introduced in Ref. \cite{farid2006evolving}.

\subsubsection{Mean vertex degree.}

In \cite{farid2006evolving}, the mean vertex degree is calculated analytically for the case of $p_{\mathrm{del}}=p_{\mathrm{dup}}=(1-\delta)=1$, and for arbitrary dimerization rate $\alpha$. The calculation is based on solving the following two-step rate equation

\begin{equation}
	\langle E_{t+2} \rangle = \langle E_t \rangle - \langle k_t \rangle - (1-\delta)\langle k_{t+1}\rangle + \alpha.
\end{equation}

Noteworthy, unlike the majority of other models, here $t \neq N$. It is both assumed that $\langle k_N \rangle \sim \langle E_N \rangle/N$, and that the mean vertex degree is stationary as $t \rightarrow \infty$ (with a numerical confirm provided in \cite{farid2006evolving}), allowing one to write

\begin{equation}
	\langle k_N \rangle \sim \alpha(N-1),
\end{equation}

which, indeed, holds in the long time limit \cite{farid2006evolving}. The authors of \cite{farid2006evolving} highlight the relevancy of dimerization in finite-sized networks for a stationary mean vertex degree.

\subsubsection{Mean number of edges.}

Similarly to the mean vertex degree, one can obtain the mean number of edges by using $\langle E \rangle = \langle k \rangle N/2$, which yields

\begin{equation}
	\langle E_N \rangle  \sim \alpha \frac{N(N-1)}{2}.
\end{equation}

Also here, the above scaling holds in the large $N$ limit.

\subsubsection{Vertex degree distribution.}

The vertex degree distribution for this model was proposed in \cite{farid2006evolving}, yet, for the general case, it was not solved analytically. The rate equation describing changes in the vertex degree distribution is a two-step rate equation. From $t \rightarrow t+1$ it reads as \cite{farid2006evolving}

\begin{equation}
N_k^{(t+1)} = N_k^{(t)} + p_{\mathrm{del}} \left[ -\frac{N_k^{(t)}}{N^{(t)}} - k \frac{N_k^{(t)}}{N^{(t)}} + (k+1)\frac{N_{k+1}^{(t)}}{N^{(t)}} \right],
\label{farid1mast}
\end{equation}

where the first loss term in squared parentheses considers the deletion of a vertex of degree $k$, the second loss term considers the deletion of an adjacent vertex of a vertex of degree $k$, and the third term (which is a gain term) considers the deletion of an adjacent vertex of a vertex of degree $k+1$.

Then, for $t+1 \rightarrow t+2$, the rate equation reads as \cite{farid2006evolving}

\begin{equation}
\begin{aligned}
	N_k^{(t+2)} = N_k^{(t+1)} + p_{\mathrm{dup}}\left\{ \frac{N_k^{(t+1)}}{N^{(t+1)}}\left[ - \alpha - (1-\delta)k\right]  + \frac{N_{k-1}^{(t+1)}}{N^{(t+1)}} \left[ \alpha + (1-\delta)(k-1)\right] + \mathcal{M}_k \right\}.
\end{aligned}
\label{farid2mast}
\end{equation}

The first term in curly brackets, a loss term, considers the duplication of a vertex with degree $k$ with dimerization (with probability $\alpha$), and an adjacent vertex of the duplicated vertex of degree $k$ which does not lose an edge (with probability $1-\delta$). The second term in curly brackets (a gain term) considers a vertex of degree $k-1$ that is duplicated with subsequent dimerization (with probability $\alpha$), and an adjacent vertex of the duplicated vertex of degree $k-1$ that does not lose an edge (with probability $1-\delta$). The last term in curly brackets, i.e., $\mathcal{M}_k$, reads as \cite{farid2006evolving}

\begin{equation}
	\begin{aligned}
\mathcal{M}_k &= (1-\alpha)\sum_{s=k}^{\infty} \frac{N_s^{(t+1)}}{N^{(t+1)}} \binom{s}{k}(1-\delta)^{s-k} \delta^{k} + \alpha \sum_{s=k-1}^{\infty} \frac{N_s^{(t+1)}}{N^{(t+1)}} \binom{s}{k-1}(1-\delta)^{s-(k-1)} \delta^{k-1},
\end{aligned}
\end{equation}

which is reminiscent of the $\mathcal{M}_k$ seen in Eq.~(\ref{Gk_ispo}) and in Eq.~(\ref{G_kcai}) with proper modifications introduced by this model. The authors of \cite{farid2006evolving} provide numerical solution to Eq.~(\ref{farid1mast})-(\ref{farid2mast}), and studied the model via finite-size scaling (see, Ref.~\cite{farid2006evolving}).

\subsubsection{Other models with deletion.}

Deletion principles in duplication models have been extended from the model in \cite{farid2006evolving}. For instance, in \cite{thornblad2015asymptotic}, deletion happens similarly as in \cite{farid2006evolving} and duplication is a full duplication as shown in Sec.~\ref{sec_fulldup}. The generic iteration ($t \rightarrow t+1$) of the full duplication model with deletion, introduced in \cite{thornblad2015asymptotic}, is

\begin{tcolorbox}[boxrule=0.8pt, colframe=white, colback=white, sharp corners]
\begin{enumerate}[leftmargin=30pt,itemsep=-0.01in]
	\item With probability $p_{\mathrm{dup}}$, a vertex $i$ is chosen uniformly at random among existing $t$ vertices and it is duplicated into vertex $i'$.
	\item With probability $p_{\mathrm{del}}$, a vertex $i$ is chosen uniformly at random among existing $t$ vertices (thus, not considering $i'$), and it is removed (with all its edges).
\end{enumerate}
\end{tcolorbox}

Removal of edges and vertices has been proposed differently from the model introduced in \cite{farid2006evolving}. Recent efforts propose both edge deletion \cite{lo2025isolated} and vertex loss \cite{zhang2024simulating} in duplication-divergence models. Also, in \cite{thornblad2015asymptotic}, deletion is combined with a full duplication model (Sec.~\ref{sec_fulldup}); similar duplication models with deletion principles have been studied in \cite{backhausz2015asymptotic,backhausz2016further}. In \cite{hamdi2014tracking}, the duplication with complete asymmetric divergence (Sec.~\ref{sec_partialdup}) was extended with deletion of a randomly chosen vertex with all its edges, and similarly, the authors of \cite{hermann2021partial} introduced a model of duplication with complete asymmetric divergence, in which a deletion rate affects each edge of the copy vertex subsequent to duplication.

\subsection{Generalizations of prior models}
\label{genrl}

The \textit{duplication-divergence-dimerization-mutation} model by Cai \textit{et al.} \cite{cai2015mean} (Sec.~\ref{DDHM}) has provided a general procedure that encompasses many of the introduced duplication (and divergence) models. In \cite{borrelli2024divergence}, an even more general procedure is introduced, which includes a new rate that the author calls the \textit{divergence asymmetry rate} $\sigma \in [0,1]$.

The divergence asymmetry rate allows to obtain a full range of structural configurations between two known limit cases: (i) the complete asymmetric divergence, and (ii) the coupled symmetric divergence. Indeed, prior models have only explored a value of the divergence asymmetry rate that is either $\sigma=1$ (complete asymmetric divergence) or $\sigma=1/2$ (coupled symmetric divergence).

Generalizing the extent of asymmetry in the divergence process yields duplication (and divergence) model graphs formed by connected components of heterogeneous size (such a structural feature is deepened in the discussion section). Furthermore, in the duplication process, the author of \cite{borrelli2024divergence} introduces via $d \in \lbrace 0,1 \rbrace$ the possibility to either consider ($d=0$) or to exclude ($d=1$) the non-interacting vertices for duplication, generalizing a number of different duplication-divergence models. Table \ref{tab1_borr} summarizes the contribution of this generalization, showing how particular values of parameters in this more general model can retrieve a number of prior known models.

\begin{table}
	\caption{The generalization in \cite{borrelli2024divergence} can be specialized to cover prior duplication (and divergence) models, as well as models such as, e.g., the random recursive tree, Barabási-Albert (as shown in \cite{ispolatov2005duplication}), urn model. Note that `any' means $d,\sigma,\delta,\alpha,\beta \in [0,1]$, while `irrelevant' means that the parameter indicated in the column header does not affect the model behavior. `Ref. Model Name' reflects the name of reference models to which this generalization can be reduced to, while `Sec.' points out (when available) to sections of this review where the model was introduced. `$G_{t_0}$' reports either an initial graph used in `Ref.', or a proposed one compatible with \cite{borrelli2024divergence}. For the Barabási-Albert model$^*$ the equivalence holds only in terms of the limiting vertex degree distribution.}
	\vspace{0.05in}
	\centering
	\begin{tabular}{@{}ccccccccc@{}}
\toprule
$d$        & $\sigma$      & $\delta$ & $\alpha$ & $\beta$ & $G_{t_0}$ & Ref. Model Name                    & Sec. & Ref. \\ \midrule
          irrelevant & irrelevant    & 0        & 0        & 0      & \initg & Full duplication model     &   \ref{sec_fulldup}   &   \cite{kumar2000stochastic}    \\
           1          & 1             & any      & 0        & 0      & \initg & Duplication-divergence   &  \ref{sec_partialdup}    &  \cite{ispolatov2005duplication}     \\
           0          & $1$ & any      & 0      & any     &  \initg & Solé \textit{et al.} model          &   \ref{sec_dupdivmut}    &   \cite{sole2002model}    \\
			0          & $1$ & any      & 0      & any     &  \initg & Kim \textit{et al.} model          &   \ref{sec_dupdivmut}    &   \cite{kim2002infinite}    \\
           0          & 1             & any      & 0      & any    & \initg & Pastor-Satorras \textit{et al.} model &  \ref{sec_dupdivmut}   &    \cite{pastor2003evolving}   \\
          irrelevant       & 1           & any      & 1        & 0       & \initgrtt & Vertex copying model                 &   \ref{sec_dupdivdim_coup}  & \cite{bhat2016densification}     \\
          0          & $1/2$ & any      & any      & 0     &  \initg & Vazquez \textit{et al.} model          &   \ref{sec_dupdivdim_coup}    &   \cite{vazquez2003modeling}    \\
          0          & $1/2$ & any      & any      & 0      & \initgsolval & Solé-Valverde model            &  \ref{sec_dupdivdim_coup}     &   \cite{sole2008spontaneous}    \\
          1          & 1             & $1^{-}$ & 0        & 0      & \initg & Barabási-Albert model$^*$         &   -   &  \cite{barabasi1999mean}     \\
			irrelevant        & irrelevant    & 1        & 1        & 0      & \initgrtt & Random recursive tree       &      - &    \cite{smythe1995survey}   \\
          irrelevant & irrelevant    & 0        & 1        & 0     & \initgpol & Eggenberger-Pólya urn    &  -    & \cite{eggenberger1923statistik} \\
          0 & 1    & 1        & 0       & any     & \initgrtt & Krapivsky-Derrida model    &  -    & \cite{krapivsky2004universal} \\
          \bottomrule
         
\end{tabular}
\label{tab1_borr}
\end{table}

\section{Discussion}

\subsection{Multifractality in time-dependent vertex degree distribution}
Multifractality is a peculiar emergent feature of duplication (and divergence) growing graph models \cite{vazquez2003growing}. We have typically considered a stationary vertex degree distribution in the thermodynamic limit, and assumed the proportion of vertices with degree $k$ as an extensive quantity. In that case, the lhs of the rate equation for the vertex degree distribution was reduced to $dN_k/dt = n_k$. Instead, when one considers a non-stationary vertex degree distribution, the lhs of the rate equation for the vertex degree distribution reads as

\begin{equation}
\frac{\partial [t n_k(t)]}{\partial t} = t \frac{\partial n_k(t)}{\partial t} + n_k(t).
\end{equation}

It is then possible to calculate $l$th-moments (i.e., $M_l(t)$) of the vertex degree distribution from its rate equation by summing over $k$ and multiplying by $k^l$, i.e., $M_l(t) = \sum_k k^l n_k(t)$. For instance, for the linear preferential-attachment model, namely the Barab{\'a}si-Albert model \cite{barabasi1999mean}, one gets

\begin{equation}
M_l(t) \propto t^{\tau^{(pa)}(l)},
\end{equation}

where the exponent $\tau^{(pa)}(l)$ is a linear function of $l$ \cite{dorogovtsev2003evolution}. This result reflects an underlying fractal distribution $n_k(t)$: a single exponent can describe its scaling behavior. Conversely, as seen in Sec.~\ref{Sec_multifract} for duplication (and divergence) models, the $l$th-moment reads as

\begin{equation}
M_l(t) \propto t^{\tau^{(dd)}(l)},
\end{equation}

with an exponent $\tau^{(dd)}(l)$ that is instead nonlinear function (see Eq.~(\ref{eq_MF1})), reflecting a \textit{multifractal} distribution $n_k(t)$. A multifractal distribution cannot be characterized by a single exponent, but it demands an infinite number of exponents, an emergent behavior found in many physical systems (e.g., see \cite{benzi1984multifractal,de1987multifractal,stanley1988multifractal,coniglio1989fractals,coniglio1989multiscaling,baity2024multifractality}, to mention only a few). Multifractality was similarly observed in sequentially growing network models that combine the preferential-attachment principle with the duplication-divergence principle, as in the model introduced in \cite{dorogovtsev2002multifractal}.

\subsection{Duplication at the origin of preferential-attachment}

The preferential-attachment principle for growing networks has attracted a wide interest from many scientific fields.\footnote{The same (or in same cases highly similar) principle has been often referred to as \textit{Zipf law} \cite{auerbach1913gesetz,zipf1942unity}, \textit{Lotka law} \cite{lotka1926frequency}, \textit{Gibrat law} \cite{gibrat1931sirey}, \textit{Yule process} \cite{willis1922some, yule1925ii}, \textit{the rich-get-richer} \cite{simon1955class}, \textit{Matthew effect} \cite{merton1968matthew}, \textit{cumulative advantage} \cite{price1976general}. For a review of mechanisms generating power-laws, see \cite{newman2005power}.} Concerning the emergence of preferential-attachment in growing networks, the duplication (and divergence) growth principle is among the possible origins of preferential-attachment in evolving networks, among other principles such as, e.g., optimization principles \cite{d2007emergence,papadopoulos2012popularity}, and the random uniform link-selection introduced in \cite{dorogovtsev2001size}.

Each of the aforementioned hypotheses are supported by showing proportionality to linear preference in the attachment rate $A_k$ of new vertices to existing vertices of growing graphs.

For the simpler case of duplication with complete asymmetric divergence (without non-interacting vertices) let one consider the variation of the number of vertices with degree $k$ (i.e., Eq. \ref{mastdeg1})

\begin{equation}
\nu \frac{dN_k}{dN} = (1-\delta) \left[ (k-1)\frac{N_{k-1}}{N} - k \frac{N_k}{N} \right] + \mathcal{M}_k.
\end{equation}

The above equation can be rewritten by defining $A_k = (1-\delta)k$

\begin{equation}
\nu \frac{dN_k}{dN} = A_{k-1}\frac{N_{k-1}}{N} - A_{k}\frac{N_{k}}{N} + \mathcal{M}_k.
\label{masteqISPO_discuss}
\end{equation}

One can immediately notice that for $\delta \rightarrow 1^{-}$ the proportionality of the attachment rate is $A_k \propto k$ and the duplication-divergence model yields linear preferential-attachment. Indeed, in the Barab{\'a}si-Albert model, $N_k$ evolves according to

\begin{equation}
\frac{dN_k}{dN} = \frac{A_{k-1}^{(pa)}}{\sum_j A_j N_j} N_{k-1} - \frac{A_{k}^{(pa)}}{\sum_j A_j N_j} N_k + \delta_{k,1},
\label{bara_discuss}
\end{equation}

with the Kronecker delta as the last term on rhs. The general preferential-attachment rate is $A_k^{(pa)} \propto k^{\epsilon}$, with $\epsilon \geq 0$, noting that $A_k^{(pa)} \sim A_k$ when $\epsilon=1$, giving the linear attachment rate of \cite{barabasi1999mean}. Indeed, it was shown in \cite{ispolatov2005duplication}, that solving (\ref{masteqISPO_discuss}) for $\delta \rightarrow 1^-$ ($p \rightarrow 0^+$) gives the same solution to (\ref{bara_discuss}) (from \cite{barabasi1999mean})---a vertex degree distribution that scales as a power-law $n_k \sim k^{-3}$.

\subsection{Percolation, connected components, and modules}

Duplication (and divergence) models can generate graphs with connected components, namely, sub-graphs with vertices connected to each other but not connected to vertices of other sub-graphs. This observation holds for the duplication-divergence with both coupled and uncoupled symmetric divergence (respectively, Sec.~\ref{sec_dupdivdim_coup}, and Sec.~\ref{sec_dupdivdim_uncoup}).

Connected components are present in a variety of networks, for instance, in the traditional Erd{\H{o}}s-R{\'e}nyi (Flory-Stockmayer) random graphs \cite{erdHos1961strength}, in some preferential-attachment network models \cite{zen2007percolation}, in general growing network models \cite{krapivsky2004universal}, and in diverse empirical networks (e.g., \cite{primario2017measuring,babvey2020content}). Except from a few cases (e.g., \cite{kim2002infinite}), there has not been a particular emphasis on sequentially growing random graph models with connected components that arise from the principle of duplication (and divergence). Nonetheless, graphs with connected components are relevant because this kind of structure relates to the problem of percolation on graphs \cite{kim2002infinite,krapivsky2004universal}, which can be of interest to the study of phenomena such as diffusion processes on networks \cite{newman2007component}. The authors of \cite{sole2008spontaneous} studied modularity in duplication and symmetric coupled divergence with dimerization (model in Sec.~\ref{sec_dupdivdim_coup}). Through a modularity maximization algorithm for community detection in graphs \cite{blondel2008fast,lancichinetti2008benchmark,fortunato2010community,fortunato202220}, the authors of \cite{sole2008spontaneous} found an increasing number of modules with increasing $\delta$ until a value ($\delta \approx 0.7$) at which the number of modules reaches a maximum and then decreases as $\delta \rightarrow 1$ from below. As mentioned in \cite{sole2008spontaneous}, when $\delta \ll 1$, the graph has a very high number of edges with all vertices being part of the largest connected component, and a high average vertex degree $\langle k \rangle$ (Eq.~(\ref{sols_sub})), thus, a large number of modules is not expected.  As $\delta$ increases, the network starts to get slowly fragmented in connected components: while the largest connected component may remain visible (depending on $\alpha$ as well), it would be made of loosely linked modules (having a lower $\langle k \rangle$). The number of modules reaches a maximum until the graph starts to get fragmented into many small connected components exhibiting a low value of modularity \cite{sole2008spontaneous}.

\subsubsection{Infinite-order percolation transition.}

The duplication-divergence with mutation model (Sec.~\ref{sec_dupdivmut}), when $\beta>0$ and $\delta=1$, has shown a connected components size distribution with an algebraic decay in the non-percolating phase, with a logarithmic correction in the percolating phase (see, Ref.~\cite{kim2002infinite}). These observations emerge as special cases of universal properties of graphs growing by attaching a new vertex to $k$ existing vertices with probability $p_k$ \cite{krapivsky2004universal}. In the special case of \cite{kim2002infinite}, $p_k$ takes the form of a Poisson distribution $p_k = (\beta^k / k!)e^{-\beta}$, which it was studied in \cite{bauer2003simple}. As pointed out in \cite{krapivsky2001organization}, such features are reminiscent of the Berezinskii--Kosterlitz--Thouless phase transition \cite{berezinskii1971destruction,kosterlitz1973ordering}, but the nature of this similarity is unknown.

\subsubsection{Connected components sizes and power-laws.}

Connected components have emerged in a generalization of duplication-divergence models \cite{borrelli2024divergence} (here introduced in Sec.~\ref{genrl}), where coupled symmetric divergence (Sec.~\ref{sec_dupdivdim_coup}) and complete asymmetric divergence (Sec.~\ref{sec_dupdivdim_uncoup}) are two limit cases of a continuous range of asymmetry in the divergence process given by the \textit{divergence asymmetry rate} $\sigma$. In particular cases, the number of connected components (of size greater than 1) scales as a power-law with the connected component size.  In \cite{borrelli2024divergence}, connected components arise in finite-sized graphs obtained from this general model with $\frac{1}{2} < \sigma <1$ (or, $0 < \sigma <\frac{1}{2}$, by symmetry); particularly, when $\sigma = \frac{1}{2}$ and $\delta \approx 0.7$, it is suggested that the connected components size distribution (for component size greater than 1) follows a power-law with exponent approximately equal to $-5/3$, an intriguing value that may remind the known -5/3 Kolmogorov isotropic turbulence that may have originally appeared in \cite{onsager1949statistical}.

\subsection{Fluctuations and lack of self-averaging}

Despite being one of the simplest model of sequentially growing graphs, the full duplication model shows the emergence of peculiar properties such as the lack of self-averaging (as shown in Sec~\ref{sec_fulldup}). Self-averaging means that fluctuations about the mean of a certain observable $x_t$ vanish in the limit of large graph order $t$, i.e., $\lim_{t \rightarrow \infty} \langle x_t^2 \rangle / \langle x_t \rangle ^2 = 1$, or equivalently, $\lim_{t \rightarrow \infty} ( \langle x_t^2 \rangle - \langle x_t \rangle ^2 ) = 0$. As argued in \cite{raval2003some}, this feature of the full duplication model could imply a lack of ergodicity. Giant fluctuations have been found in the duplication and complete asymmetric divergence model with mutation \cite{kim2002infinite} (Sec.~\ref{sec_dupdivmut}), in the so-called dense phase, in the limit of $\delta \rightarrow 0$.

\subsection{Formal mathematical results on reviewed models}

In \cite{bebek2006degree}, the model introduced in \cite{pastor2003evolving} (i.e., a duplication and complete asymmetric divergence with mutation) is studied and generalized \cite{bebek2006improved}, presenting mathematical arguments that could exclude a power-law with an exponential decay to describe the expected vertex degree distribution of the model studied in \cite{pastor2003evolving}. Yet, the model introduced in \cite{pastor2003evolving} is generalized to have a vertex degree $k\geq 1$, showing that it might exhibit a limiting power-law vertex degree distribution.

With respect to the duplication-divergence model introduced in Sec. \ref{sec_partialdup}, in \cite{li2013degree}, it is shown that for each $k\geq 0$, the fraction of vertices with degree $k$ approaches a stationary limit as the graph becomes increasingly large, answering an open question in \cite{bebek2006degree}. Also, in \cite{li2013degree}, it is demonstrated that, asymptotically, $p=1-\delta=1/2$ may be a transition point for the fraction of vertices with degree $k=0$ (non-interacting vertices) converging to 1 while, in contrast, the authors of \cite{hermann2016large} shown that this transition may occur almost surely at $p = W(1)$ (solution to $pe^p =1$) with $W(\cdot)$ the Lambert W-function; intriguingly, its approximate value was found in the partial duplication model concerning a solution to the exponent of the expected vertex degree distribution \cite{chung2003duplication}.

In \cite{knudsen2008markov}, a Markov chain approach to sequentially growing random graphs is introduced. In particular, the authors of \cite{knudsen2008markov} study the possibility to construct a time-inohomogeneous Markov chain  for a general class of random growing graphs among which the duplication and complete asymmetric divergence with dimerization (model in Sec.~\ref{DDHM} with $\beta=0$), for which conditions for ergodic or transient chain are obtained.  The authors of \cite{li2013degree} studied the model with duplication and complete asymmetric divergence without non-interacting vertices showing that $p=1-\delta=1/2$ is a transition point to obtain a limiting  vertex degree distribution, and a power-law vertex degree distribution may hold for $1/2 \leq \delta < 1$; their results show independence from the initial graph in agreement with observations made in prior work \cite{hormozdiari2007not}.

The duplication and complete asymmetric divergence (without non-interacting vertices) was formally studied in \cite{jordan2018connected}; in particular, results concerning the exponent of the power-law vertex degree distribution for $e^{-1} < p<1/2$ (see Eq.~(\ref{degdist_cases})) and  for $p < e^{-1}$ (see Eq.~(\ref{gamma_ispo})), obtained in \cite{ispolatov2005duplication}, are in agreement with \cite{jordan2018connected} and further confirmed in \cite{jacquet2020power} for $p < e^{-1}$. In \cite{frieze2020degree}, for the model with duplication and complete asymmetric divergence with mutation (Sec.~\ref{sec_dupdivmut}), exact and asymptotic results of the mean vertex degree and the degree of a fixed vertex were shown. Recent efforts consider graph symmetry as the $\text{ln}(|\text{Aut}(G)|)$  \cite{frieze2020degree}, where $|\text{Aut}(G)|$ is the cardinality of the automorphism group of a random graph $G$, i.e., the number of permutations among all possible permutations of vertices that preserve adjacencies of $G$.  Graph symmetry is intimately related to quantities such as the the mean vertex degree, the degree sequence, and the vertex degree distribution as, for instance, the number of symmetrical graphlet structures (e.g.,  \cherry, and  \diamd) is a lower bound for $|\text{Aut}(G)|$. Moreover, it has been shown that random growing graphs by preferential-attachment are asymmetric with a high probability \cite{luczak2019asymmetry} as well as traditional random graphs such as the Erd{\H{o}}s-R{\'e}nyi that exhibit $\text{ln}(|\text{Aut}(G)|)=0$, see \cite{choi2012compression}. However, for a restricted range of values of divergence rates, duplication (and divergence) graph models and some empirical networks (that may be not solely confined to the biological context) showed high extent of graph symmetry (see \cite{frieze2020degree} and \cite{sreedharan2020revisiting}). For the full duplication model (Sec.~\ref{sec_fulldup}), in \cite{turowski2020compression}, it was shown that, asymptotically, $\langle \text{ln}(|\text{Aut}(G_t)|) \rangle \sim t\text{ln}(t)$, signifying a high extent of graph symmetry. This finding has brought to the study of other graphs features such as graph compression \cite{turowski2020compression}, leading to optimal asymptotic graph compression procedures for the full duplication model, but not for other duplication (and divergence) models \cite{frieze2020degree}. In \cite{sreedharan2020revisiting}, for pairwise protein interaction networks, it is shown that graph symmetry plays a role in parameter estimation (divergence rate, and mutation rate) for the duplication and complete asymmetric divergence with mutation (model in Sec.~\ref{sec_dupdivmut}).

In \cite{frieze2024concentration}, extending results published in \cite{frieze2020degree} and \cite{frieze2021concentration} for the model with duplication and complete asymmetric divergence with mutation (Sec.~\ref{sec_dupdivmut}), it is asymptotically shown that the mean vertex degree and the maximum degree are sharply concentrated around their respective mean values. The duplication and complete asymmetric divergence with mutation model (including non-interacting vertices) has been generalized in \cite{barbour2022expected} by introducing a new parameter $q$ such that when $q=0$ one can obtain the model introduced in Sec.~\ref{sec_dupdivmut} and reproduce results of \cite{jordan2018connected}; for general $q$ values it could be possible to study cases of non-stationary expected degree distribution for any value of $p$ \cite{barbour2022expected}. For the same model (which is a model with duplication and complete asymmetric divergence with mutation), exact asymptotic results (first and second moments) of the mean vertex degree, and of the vertex degree of a fixed vertex over time, were provided in \cite{turowski2021towards}.

\subsection{Analogies across diverse models}
In \cite{evlampiev2008conservation}, it is shown that evolutionary constraints strongly related to duplication (and divergence) processes affect the network structure of protein-protein interactions without any regard to specific biological functions. Duplication (and divergence) model graphs may reproduce the structure of biological networks such as protein-protein interactions, e.g, in terms of vertex degree distribution \cite{shao2014choosing} and graphlet counts \cite{colak2009dense}. The duplication principle was also studied with models of gene transcriptional networks \cite{wagner1994evolution}, arguing that evolution of these networks should preferentially occur by single gene duplication events or entire genome duplication events \cite{zhu2013evolution}. Yet, duplication (and divergence) models are of interest beyond biological networks. In the copying model studied in \cite{bhat2016densification,lambiotte2016structural}, each added vertex is attached to a random vertex, and with probability $p$ is attached to each of its adjacent vertices. This model mimics a social network growth (the authors use the analogy of online social networks), which is exactly equivalent to the model of duplication and complete asymmetric divergence with sure dimerization events, i.e., $\alpha = 1$; $\alpha=1$ always adds an edge between vertex $i'$ and vertex $i$ at each iteration as it is also assumed in \cite{chung2006complex,li2015analysis}.  Then, divergence allows to attach (with probability $p=1-\delta$) vertex $i'$ to each adjacent vertex of $i$ (friends of $i$ in a social network), thus, the model in \cite{bhat2016densification,lambiotte2016structural}, with an arbitrary $\alpha$, resembles duplication (and divergence) models. Another extension (with additional edge deletion step) of the duplication and complete asymmetric divergence in the context of social networks is the one in \cite{hamdi2014tracking}, which also sets $\alpha=1$. Motivated by the context of social networks, a model named the \textit{friend-of-a-friend} model \cite{levens2022friend,lo2024properties} turns out to be an equivalent formulation of a duplication and complete asymmetric divergence with dimerization rate $\alpha \in [0,1]$.

\subsubsection{Mixed, content-based, directed models.}

A few prior efforts combined duplication models with preferential-attachment \cite{raval2003some,cohen2010preferential,dorogovtsev2002multifractal}.  So far, by model construction and definition, duplication models have typically considered simple undirected graphs. In \cite{krapivsky2005network}, a directed graph for a duplication model is instead introduced. In other peculiar cases, each vertex is associated with a diverse sequence of symbols. In some sense, such sequences enable the possibility of considering diverse kinds of vertices, a possibility that was typically neglected in other duplication (and divergence) models. This kind of approach led to calling them \textit{content-based network models} \cite{csengun2006content}.

\subsection{Underlying geometry of growing graphs}
In statistical physics, one can typically study either equilibrium systems, or, out of equilibrium (non-equilibrium) systems. Similarly, this distinction applies to graphs and networks. Yet, the duality of equilibrium graphs and non-equilibrium (growing) graphs has been questioned (see, \cite{krioukov2013duality}). For certain classes of random geometric graphs (e.g., causal sets in de Sitter space-time and Lorentzian spaces, the latter of interest to approach quantum gravity) and under certain conditions, the equilibrium and the non-equilibrium formulations yield two statistical ensembles of graphs, one for the equilibrium formulation and the other one for the non-equilibrium formulation, both showing, for any given graph in one ensemble, the same probability of realization that is found in the other ensemble, thus a form of equivalence of the two formulations \cite{krioukov2013duality}.

In this respect, concerning sequentially growing random graphs, e.g., preferential-attachment model graphs and plausibly duplication (and divergence) model graphs, there may not be an equivalent equilibrium formulation of these growing (non-equilibrium) graph models, as suggested in \cite{krioukov2013duality}.  Among network growth models in underlying metric spaces within the context of \textit{network geometry} \cite{boguna2021network}, to date, none of them seems to leverage the duplication (and divergence) principle, for which, published network growth models (reviewed here) are not embedded in any underlying manifold where the duplication (and divergence) principle shapes the formation of linkages among vertices. Based on results on the popularity-similarity optimization principle and the hyperbolic nature of growing tree-like graphs \cite{papadopoulos2012popularity}, non-Euclidean hidden metric spaces could be a possible approach to advance knowledge of growing graph models based on duplication (and divergence), thus, more research is expected to follow this direction.

Furthermore, the increasing attention that is being paid in network science toward higher-order networks (e.g., simplicial complexes and hypergraphs) highlights possible future research in duplication (and divergence) network growth models  \cite{battiston2020networks,bianconi2021higher,battiston2021physics,boccaletti2023structure}. Indeed, so far, only traditional pairwise edges of simple graphs have been considered in prior work reviewed here. Yet, models that were here tackled were simple undirected graphs, but approaches based on high-order graphs may be expected. Yet, this review also discussed that, based on published work on duplication (and divergence) models, there may still be phenomenological aspects to deepen in traditional pairwise edge duplication (and divergence) growing graphs, which requires further research.

\section{Conclusion}
Duplication (and divergence) models are a relevant class of sequentially growing random graph models, inspired by the biological context and considered among possible principles underlying the emergence of preferential-attachment, a widely known growing graph model that generates sparse graphs with a power-law vertex degree distribution. The substantial scientific literature available on duplication-divergence models is rich of contributions from a wide range of academic disciplines (e.g., theoretical physics, computer science, quantitative biology, mathematics).

Given the potential general character of duplication (and divergence) models, the interest in these models has begun to be extended beyond scientific fields that were the initial target audience. Indeed, on the one hand, the principle of duplication (and divergence) was inspired by a theory of evolution by gene duplication \cite{ohno1970evolution}, then, it was argued to be related to universal principles of network growth at the origin of complexity \cite{sole2020evolving}, and a `multiversal' principle as one might conjecture when considering the emergence of connected components due to non-complete divergence asymmetry. Then, while first duplication (and divergence) models were published in the late 1990s and early 2000s, the emergent complexity and the rich phenomenology of duplication (and divergence) model graphs may present a potential lack of a complete understanding, and novel synthesis efforts such as, for instance, the formulation of generalizations of prior duplication (and divergence) models, e.g., the recent one proposed in \cite{borrelli2024divergence}, may be required. On the other hand, interfacing these models with heterogeneous systems evolving through similar principles might enable the possibility to provide extensions and deepening of already published duplication (and divergence) models. 

The scope of this work was to provide a comprehensive and an introductory review of duplication (and divergence) models, cataloging and concisely reviewing mean-field results of typical quantities (e.g., the mean vertex degree, number of edges, and vertex degree distribution), with a discussion on intriguing themes for further research.

\section*{Acknowledgments}
D.B. acknowledges Prof. A. Coniglio for an inspiring seminar held at the Department of Physics, University of Naples Federico II.

\end{sloppypar}
\cleardoublepage
\phantomsection
\addcontentsline{toc}{section}{References}
\section*{References}
\vspace{-0.1in}
\bibliographystyle{unsrt}
\bibliography{biblio_models_dois.bib}

\end{document}